\documentclass[aip,jcp,amsmath,amssymb,reprint]{revtex4-2}
\usepackage[pdftex]{graphicx}
\usepackage[version=3]{mhchem}
\usepackage{rotating}
\usepackage{multirow}
\usepackage{physics}

\usepackage{amsmath}
\usepackage{bm,braket}

\newcommand{ \Sec }[1]{Sec.~\ref{sec:#1}}
\newcommand{ \Appendix }[1]{Appendix \ref{sec:#1}}

\newcommand{ \Eq   }[1]{Eq.~(\ref{#1})}
\newcommand{ \Eqs  }[2]{Eqs.~(\ref{#1}) and (\ref{#2})}

\newcommand{ \Table }[1]{Table \ref{tab:#1}}

\newcommand{ \Refs }[2]{Refs.~\onlinecite{#1} and \onlinecite{#2}}

\newcommand{ \Fig     }[1]{Fig.~\ref{fig:#1}}

\begin{document}

\title{A methodology of quantifying membrane permeability based on returning probability theory and molecular dynamics simulation}

%
\author{Yuya Matsubara}
\affiliation{Division of Chemical Engineering, Graduate School of Engineering Science, Osaka University, Toyonaka, Osaka 560-8531, Japan}
\author{Ryo Okabe}
\affiliation{Division of Chemical Engineering, Graduate School of Engineering Science, Osaka University, Toyonaka, Osaka 560-8531, Japan}
\author{Ren Masayama}
\affiliation{Division of Chemical Engineering, Graduate School of Engineering Science, Osaka University, Toyonaka, Osaka 560-8531, Japan}
\author{Nozomi Morishita Watanabe}
\affiliation{Division of Chemical Engineering, Graduate School of Engineering Science, Osaka University, Toyonaka, Osaka 560-8531, Japan}
\author{Hiroshi Umakoshi}
\affiliation{Division of Chemical Engineering, Graduate School of Engineering Science, Osaka University, Toyonaka, Osaka 560-8531, Japan}
\author{Kento Kasahara}\email[Author to whom correspondence should be addressed: ]{kasahara@cheng.es.osaka-u.ac.jp}
\affiliation{Division of Chemical Engineering, Graduate School of Engineering Science, Osaka University, Toyonaka, Osaka 560-8531, Japan}
\author{Nobuyuki Matubayasi}\email{nobuyuki@cheng.es.osaka-u.ac.jp}
\affiliation{Division of Chemical Engineering, Graduate School of Engineering Science, Osaka University, Toyonaka, Osaka 560-8531, Japan}
%
%

\begin{abstract}
We propose a theoretical approach to estimate the permeability coefficient 
of substrates (permeants) 
for crossing membranes from donor (D) phase to acceptor (A) phase 
by means of molecular dynamics (MD) simulation.
%
A fundamental aspect of our approach involves reformulating 
the returning probability (RP) theory, a rigorous bimolecular reaction theory, 
to describe permeation phenomena. 
This reformulation relies on the parallelism 
between permeation and bimolecular reaction processes.
In the present method, the permeability coefficient is represented 
in terms of the thermodynamic and kinetic quantities for the reactive (R) phase that exists 
within the inner region of membranes.
One can evaluate these quantities 
using multiple MD trajectories starting from phase R.
We apply the RP theory to the permeation of ethanol and methylamine 
at different concentrations (infinitely dilute and 1 mol\% conditions of permeants). 
Under the 1 mol\% condition, the present method yields a larger permeability
coefficient for ethanol ($0.12\pm0.01\,\mathrm{cm\,s^{-1}}$) than
for methylamine ($0.069\pm0.006~\mathrm{cm~s^{-1}}$), while the values of the permeability 
coefficient are satisfactorily close to those 
obtained from the brute-force MD simulations {[}$0.18\pm0.03\,\mathrm{cm\,s^{-1}}$
and $0.052\pm0.005\,\mathrm{cm\,s^{-1}}$ for ethanol and methylamine,
respectively{]}. Moreover, upon analyzing the thermodynamic and kinetic
contributions to the permeability, we clarify that a higher concentration
dependency of permeability for ethanol, as compared to methylamine,
arises from the sensitive nature of ethanol's free-energy barrier
within the inner region of the membrane against ethanol concentration.

\end{abstract}

\maketitle

\section{Introduction\label{sec:intro}}
Permeation of substrates (permeants) through cell membranes is a fundamental process for biological systems. 
Most permeants, including drug molecules, enter a cell with passive permeation driven by the concentration gradient of permeants between the donor and acceptor phases. 
Hence, the permeability coefficient that characterizes the efficiency of passive permeation is a valuable indicator for drug delivery. 
The permeability coefficient is experimentally measured through different assays, such as the parallel artificial membrane permeability assay (PAMPA)\cite{kansy1998physicochemical,avdeef2005rise,avdeef2007pampa} and the carcinoma colorectal cell-based (CaCo-2) assay.\cite{artursson2001caco,van2005caco} 
The sophisticated spectroscopy techniques are also useful for quantitative and real-time analysis of membrane permeation.\cite{sharifian2021recent} 
Since the permeation process is governed by such factors as the solubility 
and mobility of permeants in a membrane, 
the theoretical and computational approaches in the atomistic detail 
have been recognized as promising for realizing systematic analysis.
\cite{swift2013back,venable2019molecular,chipot2023predictions,gomes2023recent}

Molecular dynamics (MD) simulation is the most popular method to elucidate the detailed mechanisms of the permeation process from a theoretical point of view. 
The inhomogeneous solubility-diffusion (ISD) model\cite{diamond1974interpretation,marrink1994simulation} incorporating MD simulations has played a central role in analyzing the permeation processes.\cite{awoonor2016molecular,venable2019molecular} 
In this model, the permeability coefficient is expressed using the free energy profile and position-dependent diffusion coefficient\cite{berne1988classical, woolf1994conformational} along the reaction coordinate for the permeation process. 
The ISD model has prompted researchers to develop methodologies for efficiently computing the position-dependent diffusion coefficient from MD simulations.\cite{hummer2005position,comer2013calculating,ghysels2017position,nagai2020position} 
This model is based on the Smoluchowski equation, realizing the simple treatment of permeation processes. 
However, the difficulty arises when the permeant shows the subdiffusive motion inside the membrane in the long-time limit.\cite{munguira2016glasslike, chipot2016subdiffusion} 
In such a case, employing the Smoluchowski equation is inappropriate. 
Recently, alternative MD-based approaches have been developed. 
Thanks to the recent advances in computers, the methodologies based on the direct observation of the permeation events are available for the permeants exhibiting the fast permeation kinetics.\cite{ghysels2019permeability,venable2019molecular,kramer2020membrane, davoudi2021sampling} 
The flux-based counting (FBC) and transition-based counting (TBC) methods enable us to estimate the permeability coefficient reliably without resorting to any theoretical models. 
Furthermore, the kinetic models such as the Markov state model (MSM) for permeation constructed with the enhanced sampling methods yielded the permeability coefficients qualitatively correlated with the experimental measurements.\cite{ghaemi2012novel,cardenas2014modeling, badaoui2018calculating,harada2022free,mitsuta2022calculation}
A methodology to calculate the permeability coefficients without assuming the Markovianity of the permeation dynamics was also developed using the equilibrium path ensemble.\cite{ghysels2021exact}

The theoretical framework for molecular binding kinetics such as protein-ligand binding could be useful for elucidating the permeation processes. 
Votapka and Amaro derived the theoretical relationship between the permeability coefficient and mean first passage time (MFPT), that was related to the rate constant in the protein-ligand binding\cite{szabo1980first, bowman2013introduction} and was defined in the context of permeation as the average time for the permeant to arrive at the acceptor phase from the donor phase for the first time.\cite{Votapka_2016} 
Since many kinetic theories have been developed to compute the MFPT, 
the relationship between the MFPT and permeability coefficient is useful to develop new methodologies for elucidating the permeation processes based on these theories.
%
They also derived the theoretical expression of the permeability coefficient using crossing probability that is suitable for the milestoning-based method.\cite{Votapka2015, Votapka_2017}

Recently, we developed an MD-based methodology for elucidating molecular binding kinetics\cite{Kasahara_2021,kasahara2023elucidating} using the returning probability (RP) theory,\cite{kim2009rigorous} a rigorous diffusion-influenced reaction (DIR) theory. 
The RP theory is based on the Liouville equation of the phase space densities with the reaction sink term that describes the reaction (binding) probability on the reactive state existing in a binding process between the dissociated and bound states. 
The perturbative expansion of the reactant distribution yields the theoretical expression of the binding rate constants suitable for MD simulations. 
It has been demonstrated that the RP theory gives the binding rate constants for the inclusion\cite{Kasahara_2021} and protein-ligand binding\cite{kasahara2023elucidating} systems consistent with those evaluated through the long-time MD simulations. 
Thanks to the analytical nature of the RP theory, furthermore, the binding kinetics is characterized in terms of the thermodynamic and kinetic properties of the reactive state. 
Hence, applying this theory gives the physicochemical insights into the binding kinetics in addition to the binding rate constants. 
Accordingly, establishing the framework for analyzing the permeation processes with the RP theory could be useful to unveil the detailed permeation mechanism.

In the present study, we develop an MD-based methodology for quantifying the permeability coefficient for membrane systems. 
We first derive the exact relationship between the permeability coefficient and the permeant distribution function at unsteady state which is similar to that between the binding rate constant and reactant distribution function for the binding systems. 
Then, by employing the perturbative expansion technique utilized in the RP theory, the tractable expression of the permeability coefficient at steady state is derived. 
In this expression, the coefficient is represented in terms of the thermodynamic and kinetic properties of the permeants inside the membrane. 
Thus, by computing these properties with MD simulations, the estimation of the permeability coefficients is realized.
%

We apply the present method to the permeation processes of ethanol and methylamine through the lipid bilayer composed of 1-palmitoyl-2-oleoyl-\textit{sn}-glycero-3-phosphocholin (POPC). 
Recently, Ghorbani \textit{et al.} investigated ethanol permeation at different concentrations using the ISD model and counting-based methods such as FBC and TBC methods.\cite{Ghorbani_2020} 
We also employ the TBC method for both ethanol and methylamine under the 1 mol\% condition to test the validity of the present method.
%

\section{Theory\label{sec:theory}}

\subsection{Permeability coefficient}
\begin{figure}[t]
  \centering
  \includegraphics[width=0.9\linewidth]{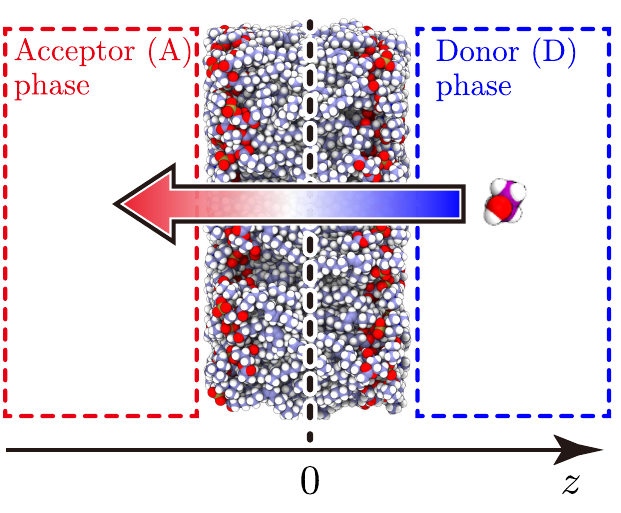}
  \caption{Membrane permeation system. $z$-direction is normal to the membrane surface and $z=0$ coincides with the center of mass (CoM) for the membrane.\label{fig:model_description}}
\end{figure}
We briefly introduce the definition of the permeation coefficient.
Let us consider a planar membrane system in which the donor (D) and acceptor (A) solution phases for permeants are separated by a lipid membrane (\Fig{model_description}).   
The concentrations of the permeants for phases D and A are defined as $c_{\mathrm{D}}$ and 0, respectively. 
According to the Fick's law, the flux across a membrane at the steady state, $J_{\mathrm{ss}}$,
is proportional to the concentration gradient of the permeants, $\Delta c = c_{\mathrm{D}}$, as follows:
\begin{align}
 J_{\mathrm{ss}} = c_{\mathrm{D}}\mathcal{P}_{\mathrm{ss}}. \label{J_ss} 
\end{align}
Here, $\mathcal{P}_{\mathrm{ss}}$ is the permeability coefficient and the subscript ss means steady state.
The generalization of \Eq{J_ss} to the unsteady state 
is possible by considering the time-dependent flux and permeability coefficient as
\begin{align}
 J\left(t\right) = c_{\mathrm{D}}\mathcal{P}\left(t\right). \label{Jt_1}
\end{align}
$J\left(t\right)$ is defined as the number of the permeants 
moving to phase A per unit area and time. 
Thus, $J\left(t\right)$ can be described by
\begin{align}
 J\left(t\right) & =-\dfrac{1}{\sigma}\dfrac{d}{dt}N\left(t\right), \label{Jt_2}
\end{align}
where $N\left(t\right)$ is the number of the permeants that are present 
in a membrane or in phase D at time $t$, and $\sigma$ is the cross-sectional area of a membrane.
Note that $J\left(t\right)$ and $\mathcal{P}\left(t\right)$ at $t\to \infty$ coincide with $J_{\mathrm{ss}}$ and $\mathcal{P}_{\mathrm{ss}}$, respectively. 
\subsection{Returning probability (RP) theory for membrane permeation}
The returning probability (RP) theory provides a theoretical fundament to 
elucidate host-guest binding phenomena such as protein-ligand binding 
based on the Liouville equation of the phase space density
with the reaction sink term that describes the insertion  
of the guest to a binding site of the host.
The theoretical expression of the binding rate constant derived by the RP theory 
is applicable to the various types of binding processes.\cite{Kasahara_2021, kasahara2023elucidating} 
In this subsection, we show that the RP theory for membrane permeation can be constructed in a similar way as for host-guest binding. 
In the RP treatment of host-guest systems, 
the binding proceeds from the dissociated (initial) state 
through the reactive (intermediate) state to the bound (final) state. 
For the membrane permeation, the D and A phases are the initial and final states, respectively, 
and a ``reactive phase'' is introduced as the intermediate configurations 
within the membrane which the permeant is to pass through.

We consider a membrane system that contains $N_{0}$ permeant molecules
and is in equilibrium at time $t\leq 0$.
The center of mass (CoM) for the $i$th permeant molecule at time $t$ is defined as ${\bf r}_{i}=\left(x_{i}\left(t\right), y_{i}\left(t\right), z_{i}\left(t\right)\right)$, where $z$-direction is normal to the membrane surface and $z=0$ coincides with the CoM for the membrane.
Then, we assume that the permeation process of the $i$th permeant molecule
can be described using the reaction coordinates composed of $z_{i}\left(t\right)$ and residual part, $\bm{\Lambda}_{i}\left(t\right)$.
$\bm{\Lambda}_{i}\left(t\right)$ represents the orientation and intramolecular degrees of freedom for the $i$th permeant molecule.
Let us define the phase space density, $\Psi_{i}\left(\bm{\Gamma},t\right)$, 
that the $i$th permeant molecule does not experience the transition to phase A and  
the phase space coordinate of the system is $\bm{\Gamma}$ at time $t$.
We also define the reactive (R) phase located around the free energy barrier
in the membrane, $\bm{\Upsilon}$.
The phase R is called so in analogy with the reactive state in the host-guest binding process.\cite{kasahara2023elucidating}
We further assume that the permeants
in $\bm{\Upsilon}$ move to phase A at a certain frequency
represented by the first-order rate constant, $k_{\mathrm{R\to A}}$.

By introducing the following reaction sink function
\begin{align}
S\left(z_{i}\left(t\right),\bm{\Lambda}_{i}\left(t\right)\right) & =\begin{cases}
k_{\mathrm{R\to A}}, & \left(z_{i}\left(t\right), \bm{\Lambda}_{i}\left(t\right)\right)\in\bm{\Upsilon},\\
0, & \left(z_{i}\left(t\right),\bm{\Lambda}_{i}\left(t\right)\right)\notin\bm{\Upsilon},
\end{cases}\label{sink_definition}
\end{align}
the differential equation for $\Psi_{i}\left(\bm{\Gamma},t\right)$
can be expressed as
\begin{align}
\dfrac{\partial}{\partial t}\Psi_{i}\left(\bm{\Gamma},t\right) & =-\mathcal{L}\Psi_{i}\left(\bm{\Gamma},t\right)-S\left(z_{i}\left(t\right),\bm{\Lambda}_{i}\left(t\right)\right)\Psi_{i}\left(\bm{\Gamma},t\right). \label{dPsi}
\end{align}
Here, $\mathcal{L}$ is the Liouville operator of the system, and
the second term of the right-hand side of \Eq{dPsi} represents the decrease
of the probability densities due to the transition of the $i$th permeant molecule
to phase A.
$\Psi_{i}\left(\bm{\Gamma},t\right)$ is normalized as
\begin{align}
\sum_{i=1}^{N_{0}}\int d\bm{\Gamma}\,\Psi_{i}\left(\bm{\Gamma},t\right) & =N\left(t\right).
\end{align}
Thus, performing the integration of \Eq{dPsi} over $\bm{\Gamma}$ and summation against the permeant molecules leads to  
\begin{align}
\dfrac{d}{dt}N\left(t\right) & =\left(\dfrac{d}{dt}N\left(t\right)\right)_{\mathrm{NR}} \notag \\
& \quad -\sum_{i=1}^{N_{0}}\int d\bm{\Gamma}\,S\left(z_{i}\left(t\right),\bm{\Lambda}_{i}\left(t\right)\right)\Psi_{i}\left(\bm{\Gamma},t\right), \label{dN_1}
\end{align}
where $(dN\left(t\right)/dt)_{\mathrm{NR}}$ is the time derivative of $N\left(t\right)$ for the hypothetical non-permeable system, in which the transition events of the permeants to phase A are absent, defined as
\begin{align}
\left(\dfrac{d}{dt}N\left(t\right)\right)_{\mathrm{NR}} & =-\sum_{i=1}^{N_{0}}\int d\bm{\Gamma}\,\mathcal{L}\Psi_{i}\left(\bm{\Gamma},t\right),
\end{align} 
and this term vanishes due to the conservation of the number of molecules.
We introduce the nonequilibrium distribution function
\begin{align}
g\left({\bf r}, \bm{\Lambda},t\right) & =\dfrac{1}{c_{\mathrm{D}}}\sum_{i=1}^{N_{0}}\int d\bm{\Gamma}\,\delta \left({\bf r} - {\bf r}_{i}\left(t\right)\right) \notag \\
& \quad \times \delta\left(\bm{\Lambda}-\bm{\Lambda}_{i}\left(t\right)\right)\Psi_{i}\left(\bm{\Gamma},t\right). \label{g_rl}
\end{align}
From \Eqs{dN_1}{g_rl}, the following equation is obtained. 
\begin{align}
\dfrac{d}{dt}N\left(t\right) & =-c_{\mathrm{D}}\int d{\bf r} \int d\bm{\Lambda}\,S\left(z,\bm{\Lambda}\right)g\left({\bf r},\bm{\Lambda},t\right). \label{dN_2} 
\end{align}
Substitution of \Eqs{Jt_2}{dN_2} into \Eq{Jt_1} gives the theoretical expression of $\mathcal{P}\left(t\right)$ as
\begin{align}
\mathcal{P}\left(t\right) & =\dfrac{1}{\sigma}\int d{\bf r} \int d\bm{\Lambda}\,S\left(z,\bm{\Lambda}\right)g\left({\bf r},\bm{\Lambda},t\right). \label{P_exact}
\end{align} 
The above expression of $\mathcal{P}\left(t\right)$ is parallel to that of the rate coefficient of molecular binding based on the RP theory.\cite{kim2009rigorous,Kasahara_2021}
In the RP theory for molecular binding, the rate coefficient of binding is exactly expressed as the integration 
of the nonequilibrium distribution function of guest molecules multiplied by the reaction sink 
function over the reaction coordinate. 
The time dependence of the nonequilibrium distribution is governed by the Liouville equation with the reaction term leading to the time evolution of $g\left({\bf r},\bm{\Lambda},t\right)$ (\Eq{g_rl}).
Thus, the perturbative-expansion technique employed in the RP theory can be adopted to derive a tractable expression of $\mathcal{P}\left(t\right)$ as the Laplace transform ($t\to s$), $\hat{\mathcal{P}}\left(s\right)$, from \Eq{P_exact}. 
The derivation is found in \Appendix{Derive_Pss}.

The resultant expression of $\hat{\mathcal{P}}\left(s\right)$ is given by
\begin{align}
s\hat{\mathcal{P}}\left(s\right) & =\dfrac{1}{\sigma}\int d{\bf r} \int d\bm{\Lambda}\,S\left(z,\bm{\Lambda}\right)g_{\mathrm{eq}}\left({\bf r},\bm{\Lambda}\right) \notag \\
 & \quad \times \left(1+k_{\mathrm{R\to A}}\hat{P}_{\mathrm{RET}}\left(s\right)\right)^{-1}.
\label{sP(s)}
\end{align}
Here, $g_{\mathrm{eq}}\left({\bf r}, \bm{\Lambda}\right)$ is the equilibrium part of $g\left({\bf r},\bm{\Lambda},t\right)$, defined as
\begin{align}
g_{\mathrm{eq}}\left({\bf r},\bm{\Lambda}\right) & =\dfrac{1}{c_{\mathrm{D}}}\sum_{i=1}^{N_{0}}\int d\bm{\Gamma}\,\delta\left({\bf r}-{\bf r}_{i}\right)\delta\left(\bm{\Lambda}-\bm{\Lambda}_{i}\right)\Psi_{\mathrm{eq}}\left(\bm{\Gamma}\right),
\label{geq_main}
\end{align}
where $\Psi_{\mathrm{eq}}\left(\bm{\Gamma}\right)$ is the phase space density at the equilibrium state.
$\hat{P}_{\mathrm{RET}}\left(s\right)$ is the Laplace transform of the returning probability, $P_{\mathrm{RET}}\left(t\right)$, defined as
\begin{align}
P_{\mathrm{RET}}\left(t\right) & =\dfrac{\displaystyle\sum_{i=1}^{N_{0}}\braket{\Theta\left(z_{i}\left(t\right), \bm{\Lambda}_{i}\left(t\right)\right)\Theta\left(z_{i}\left(0\right),\bm{\Lambda}_{i}\left(0\right)\right)}}{\displaystyle\sum_{i=1}^{N_{0}}\braket{\Theta\left(z_{i}\left(0\right), \bm{\Lambda}_{i}\left(0\right)\right)}},
\label{pret_definition}
\end{align}
where $\Theta\left(z_{i}\left(t\right), \bm{\Lambda}_{i}(t)\right)$ is the characteristic function for phase R given by
\begin{align}
\Theta\left(z_{i}\left(t\right),\bm{\Lambda}_{i}\left(t\right)\right) & =\begin{cases}
1, & \left(z_{i}\left(t\right), \bm{\Lambda}_{i}\left(t\right)\right)\in\bm{\Upsilon},\\
0, & \left(z_{i}\left(t\right), \bm{\Lambda}_{i}\left(t\right)\right)\notin\bm{\Upsilon}.
\end{cases}
\end{align}
$P_{\mathrm{RET}}\left(t\right)$ is the conditional probability of finding a permeant in $\bm{\Upsilon}$ at $t=t$ when that molecule was in $\bm{\Upsilon}$ at $t=0$. 
Owing to the final value theorem of Laplace transform, $s\hat{\mathcal{P}}\left(s\right) \xrightarrow[]{s\to 0} \mathcal{P}_{\mathrm{ss}}$, one can obtain the following expression of $\mathcal{P}_{\mathrm{ss}}$ from \Eq{sP(s)}.
\begin{align}
\mathcal{P}_{\mathrm{ss}} & =\dfrac{k_{\mathrm{R\to A}}}{\sigma}\int_{\bm{\Upsilon}} d{\bf r}\,K\left({\bf r}\right)\left(1+k_{\mathrm{R\to A}}\int_{0}^{\infty}dt\,P_{\mathrm{RET}}\left(t\right)\right)^{-1},
\label{Pss_0}
\end{align}
since $S\left(z,\bm{\Lambda}\right) = k_{\mathrm{R\to A}}\Theta\left(z,\bm{\Lambda}\right)$. 
Here, $K\left({\bf r}\right)$ is defined as
\begin{align}
K\left({\bf r}\right) & =\int d\bm{\Lambda}\,\Theta\left(z,\bm{\Lambda}\right)g_{\mathrm{eq}}\left({\bf r},\bm{\Lambda}\right).
\label{K} 
\end{align} 
By defining the concentration of the permeants in phase R at ${\bf r}$ as 
\begin{align}
c_{\mathrm{R}}\left({\bf r}\right) & =\sum_{i=1}^{N_{0}}\int d\bm{\Lambda}\,\Theta\left(z, \bm{\Lambda}\right)\int d\bm{\Gamma}\, \notag \\
& \quad \times \delta\left({\bf r}-{\bf r}_{i}\right)\delta\left(\bm{\Lambda}-\bm{\Lambda}_{i}\right)\Psi_{\mathrm{eq}}\left(\bm{\Gamma}\right),
 \label{cR}
\end{align}
$K\left({\bf r}\right)$ is expressed using Eqs. \eqref{geq_main}, \eqref{K}, and  \eqref{cR} as 
\begin{align}
  K\left({\bf r}\right) = \dfrac{c_{\mathrm{R}}\left({\bf r}\right)}{c_{\mathrm{D}}}. \label{Kstar_conc}
\end{align}
\Eq{Kstar_conc} indicates that $K\left({\bf r}\right)$ is the equilibrium constant between the position ${\bf r}$ in phase R and phase D (donor solution phase).
Since the planar membrane is uniform along with $x$- and $y$-directions, 
the concentration of the permeants only depends on $z$, i.e., $c_{\mathrm{R}}\left({\bf r}\right) = c_{\mathrm{R}}\left(z\right)$ and $K\left({\bf r}\right) = K\left(z\right)$.
Hence, \Eq{Pss_0} can be rewritten by performing the integration along the $x$- and $y$- directions as
\begin{align}
\mathcal{P}_{\mathrm{ss}} & =k_{\mathrm{R\to A}}K^{*}\left(1+k_{\mathrm{R\to A}}\int_{0}^{\infty}dt\,P_{\mathrm{RET}}\left(t\right)\right)^{-1}, \label{Pss_final}
\end{align}
where
\begin{align}
 K^{*} = \int_{\bm{\Upsilon}}dz\,K\left(z\right). \label{Kstar} 
\end{align}
%
%
%
%
%
%
%
\subsection{Theoretical expression of $K\left(z\right)$ using solvation free energies}
In this subsection, we describe an efficient scheme of calculating $K\left(z\right)$ based on the solvation free energies of the permeant molecule (solute).
The solvation free energies associated with the solvation process of the solute 
in phase D and with the solvation process at position $z$ 
in phase R are denoted as $\Delta \mu_{\mathrm{D}}$ and $\Delta \mu_{\mathrm{R}}\left(z\right)$, respectively.
According to the equilibrium condition between the two different phases, $K\left(z\right)$ can be expressed as 
\begin{align}
K\left(z\right) & =e^{-\beta\Delta G\left(z\right)},
\end{align}
where $\beta$ is the inverse temperature and $\Delta G\left(z\right)$ is the free energy profile along the $z$-direction defined as
\begin{align}
\Delta G\left(z\right) & =\Delta\mu_{\mathrm{R}}\left(z\right)-\Delta\mu_{\mathrm{D}}.
\label{Delta_G}
\end{align}
Note that $\Delta G\left(z\right)$ is equivalent to the potential of mean force (PMF) given by
\begin{align}
 \Delta G_{\mathrm{PMF}}\left(z\right) = -\dfrac{1}{\beta}\log \dfrac{c_{\mathrm{R}}\left(z\right)}{c_{\mathrm{D}}}.
 \label{dGz_PMF} 
\end{align}  

Both $\Delta \mu_{\mathrm{D}}$ and $\Delta \mu_{\mathrm{R}}\left(z\right)$ can be represented in terms of the configurational integrals. 
In the case of phase D, the solute is surrounded by only the solvent molecules (water). 
Let us define the full coordinates of the solute and the set of full coordinates of the water molecules as ${\bf x}_{\mathrm{U}}$ and ${\bf X}_{\mathrm{V}}$, respectively. 
We also express the intramolecular energy of the solute, the total potential of the solvents, and the interaction 
between the solute and solvents as $U_{\mathrm{U}}\left({\bf x}_{\mathrm{U}}\right)$, $U_{\mathrm{V}}\left({\bf X}_{\mathrm{V}}\right)$, and $U_{\mathrm{UV}}\left({\bf x}_{\mathrm{U}},{\bf X}_{\mathrm{V}}\right)$, respectively.
Then, $\Delta \mu_{\mathrm{D}}$ is given by 
\begin{align}
\Delta\mu_{\mathrm{D}} & =-\dfrac{1}{\beta}\log\dfrac{\displaystyle \int d{\bf x}_{\mathrm{U}}\int d{\bf X}_{\mathrm{V}}\,e^{-\beta\mathcal{V}_{\mathrm{D}}^{\mathrm{sol}}\left({\bf x}_{\mathrm{U}},{\bf X}_{\mathrm{V}}\right)}}{\displaystyle \int d{\bf x}_{\mathrm{U}}\int d{\bf X}_{\mathrm{V}}\,e^{-\beta\mathcal{V}_{\mathrm{D}}^{\mathrm{ref}}\left({\bf x}_{\mathrm{U}},{\bf X}_{\mathrm{V}}\right)}},
\label{eq:delta_mu_bulk}
\end{align}
where $\mathcal{V}^{\mathrm{ref}}_{\mathrm{D}}\left({\bf x}_{\mathrm{U}},{\bf X}_{\mathrm{V}}\right)$ 
and $\mathcal{V}^{\mathrm{sol}}_{\mathrm{D}}\left({\bf x}_{\mathrm{U}},{\bf X}_{\mathrm{V}}\right)$ are respectively the total potentials of the reference solvent and solution systems defined as 
\begin{align}
\mathcal{V}_{\mathrm{D}}^{\mathrm{ref}}\left({\bf x}_{\mathrm{U}},{\bf X}_{\mathrm{V}}\right) & =U_{\mathrm{U}}\left({\bf x}_{\mathrm{U}}\right)+U_{\mathrm{V}}\left({\bf X}_{\mathrm{V}}\right), \label{Uref}\\
\mathcal{V}_{\mathrm{D}}^{\mathrm{sol}}\left({\bf x}_{\mathrm{U}},{\bf X}_{\mathrm{V}}\right) & =U_{\mathrm{U}}\left({\bf x}_{\mathrm{U}}\right)+U_{\mathrm{UV}}\left({\bf x}_{\mathrm{U}},{\bf X}_{\mathrm{V}}\right)+U_{\mathrm{V}}\left({\bf X}_{\mathrm{V}}\right). \label{Usol}
\end{align}
In the reference solvent, the interaction between the solute and solvent molecules is absent.
As for phase R, both the solvent molecules (water) and membrane are relevant with the solvation thermodynamics of the solute.
Even in the presence of the membrane, $\Delta \mu_{\mathrm{R}}\left(z\right)$ can be expressed in a similar way by regarding the membrane as part of the solvent mixture and considering the conditional ensemble average using $z$ and $\bm{\Lambda}$.
We denote the solvent mixture as $\mathrm{V}^{\prime}$, and the set of full coordinates of the solvent molecules, total potential of the solvents, and the interaction of the solute with the solvents as ${\bf X}_{\mathrm{V}^{\prime}}$, $U_{\mathrm{V}^{\prime}}\left({\bf X}_{\mathrm{V}^{\prime}}\right)$, and $U_{\mathrm{UV}}\left({\bf x}_{\mathrm{U}},{\bf X}_{\mathrm{V}^{\prime}}\right)$, respectively.
The total potentials of the reference solvent and solution systems for phase R, 
$\mathcal{V}_{\mathrm{R}}^{\mathrm{ref}^{\prime}}\left({\bf x}_{\mathrm{U}},{\bf X}_{\mathrm{V}^{\prime}}\right)$ and 
$\mathcal{V}_{\mathrm{R}}^{\mathrm{sol}^{\prime}}\left({\bf x}_{\mathrm{U}},{\bf X}_{\mathrm{V}^{\prime}}\right)$ are respectively defined as the right hand sides of \Eqs{Uref}{Usol} in which $\mathrm{V}$ involved in the subscripts is replaced with $\mathrm{V}^{\prime}$.
The theoretical expression of $\Delta \mu_{\mathrm{R}}\left(z\right)$ is given by 
\begin{align}
 & \Delta\mu_{\mathrm{R}}\left(z\right)\notag\\
 & =-\dfrac{1}{\beta}\log\dfrac{\displaystyle \int d{\bf x}_{\mathrm{U}}\int d{\bf X}_{\mathrm{V}^{\prime}}\,\Phi_{z}\left(z_{\mathrm{U}},\bm{\Lambda}_{\mathrm{U}}\right)e^{-\beta\mathcal{V}_{\mathrm{R}}^{\mathrm{sol}^{\prime}}\left({\bf x}_{\mathrm{U}},{\bf X}_{\mathrm{V}^{\prime}}\right)}}{\displaystyle \int d{\bf x}_{\mathrm{U}}\int d{\bf X}_{\mathrm{V}^{\prime}}\,\Phi_{z}\left(z_{\mathrm{U}},\bm{\Lambda}_{\mathrm{U}}\right)e^{-\beta\mathcal{V}_{\mathrm{R}}^{\mathrm{ref}^{\prime}}\left({\bf x}_{\mathrm{U}},{\bf X}_{\mathrm{V}^{\prime}}\right)}},
\label{dmu_R_0}
\end{align}
where $z_{\mathrm{U}}$ and $\bm{\Lambda}_{\mathrm{U}}$ are the $z$-coordinate of the center of mass (CoM) for the solute and residual part of the reaction coordinate, respectively, and $\Phi_{z}\left(z_{\mathrm{U}}\right)$ is the characteristic function for region between $z$ and $z+\Delta z$ in phase R. 
At this point, the width along $z$ is set to be finite (but small) 
to formulate equations that are used in numerical calculations in practice.
By introducing
\begin{align}
\theta_{z}\left(z_{\mathrm{U}}\right) & =\begin{cases}
1, & z\leq z_{\mathrm{U}}<z+\Delta z,\\
0, & \mathrm{otherwise,}
\end{cases} 
\end{align}
$\Phi_{z}\left(z_{\mathrm{U}},\bm{\Lambda}_{\mathrm{U}}\right)$ can be expressed as
\begin{align}
\Phi_{z}\left(z_{\mathrm{U}},\bm{\Lambda}_{\mathrm{U}}\right) & =\Theta\left(z_{\mathrm{U}},\bm{\Lambda}_{\mathrm{U}}\right)\theta_{z}\left(z_{\mathrm{U}}\right).
\end{align}

The methodologies of computing the free energy
such as the free energy perturbation (FEP)\cite{zwanzig1954high}, thermodynamic integral (TI)\cite{kirkwood1935statistical}, 
and Bennett acceptance ratio (BAR)\cite{bennett1976efficient} can be used to obtain $\Delta \mu_{\mathrm{D}}$ 
and $\Delta \mu_{\mathrm{R}}\left(z\right)$.
%
%
In the case of $\Delta \mu_{\mathrm{R}}\left(z\right)$, the free energy calculation is required for each region defined with $z$.
To avoid the repeated free energy calculations, we adopt the scheme of calculating the profile of $\Delta \mu_{\mathrm{R}}\left(z\right)$ using the local distribution of the solute and solvation free energy for an arbitrarily defined state $\mathcal{S}$, $\Delta \mu_{\mathcal{S}}$.\cite{kasahara2023elucidating}
To simplify the notation, we introduce the ensemble averages in the reference solvent and solution systems for phase R, respectively, as  
%
%
%
\begin{align}
\braket{\cdots}_{\mathrm{ref}^{\prime}} & =\dfrac{1}{Z_{\mathrm{ref}^{\prime}}}\int d{\bf x}_{\mathrm{U}}\int d{\bf X}_{\mathrm{V}^{\prime}}\,\left(\cdots\right)e^{-\beta\mathcal{V}_{\mathrm{R}}^{\mathrm{ref}^{\prime}}\left({\bf x}_{\mathrm{U}},{\bf X}_{\mathrm{V}^{\prime}}\right)}, \\
\braket{\cdots}_{\mathrm{sol}^{\prime}} & =\dfrac{1}{Z_{\mathrm{sol}^{\prime}}}\int d{\bf x}_{\mathrm{U}}\int d{\bf X}_{\mathrm{V}^{\prime}}\,\left(\cdots\right)e^{-\beta\mathcal{V}_{\mathrm{R}}^{\mathrm{sol}^{\prime}}\left({\bf x}_{\mathrm{U}},{\bf X}_{\mathrm{V}^{\prime}}\right)},
\end{align}
where $Z_{\mathrm{ref}^{\prime}}$ and $Z_{\mathrm{sol}^{\prime}}$ are the configurational integrals for the reference solvent and solution systems, respectively, defined as
\begin{align}
Z_{\mathrm{ref}^{\prime}} &= \int d{\bf x}_{\mathrm{U}}\int d{\bf X}_{\mathrm{V}^{\prime}}\,e^{-\beta\mathcal{V}_{\mathrm{R}}^{\mathrm{ref}^{\prime}}\left({\bf x}_{\mathrm{U}},{\bf X}_{\mathrm{V}^{\prime}}\right)}, \\
Z_{\mathrm{sol}^{\prime}} &= \int d{\bf x}_{\mathrm{U}}\int d{\bf X}_{\mathrm{V}^{\prime}}\,e^{-\beta\mathcal{V}_{\mathrm{R}}^{\mathrm{sol}^{\prime}}\left({\bf x}_{\mathrm{U}},{\bf X}_{\mathrm{V}^{\prime}}\right)}.
\end{align}
Then, \Eq{dmu_R_0} is rewritten as 
\begin{align}
\Delta\mu_{\mathrm{R}}\left(z\right) & =-\dfrac{1}{\beta}\log\dfrac{\braket{\Phi_{z}\left(z_{\mathrm{U}},\bm{\Lambda}_{\mathrm{U}}\right)}_{\mathrm{sol}^{\prime}}}{\braket{\Phi_{z}\left(z_{\mathrm{U}},\bm{\Lambda}_{\mathrm{U}}\right)}_{\mathrm{ref}^{\prime}}}\dfrac{Z_{\mathrm{sol}^{\prime}}}{Z_{\mathrm{ref}^{\prime}}}.
\end{align}
If we define the characteristic function corresponding to state $\mathcal{S}$ as $\theta_{\mathcal{S}}\left(z,\bm{\Lambda}\right)$, 
$\Delta\mu_{\mathcal{S}}$ is also expressed as
\begin{align}
\Delta\mu_{\mathcal{S}} & =-\dfrac{1}{\beta}\log\dfrac{\braket{\theta_{\mathcal{S}}\left(z_{\mathrm{U}},\bm{\Lambda}_{\mathrm{U}}\right)}_{\mathrm{sol}^{\prime}}}{\braket{\theta_{\mathcal{S}}\left(z_{\mathrm{U}},\bm{\Lambda}_{\mathrm{U}}\right)}_{\mathrm{ref}^{\prime}}}\dfrac{Z_{\mathrm{sol}^{\prime}}}{Z_{\mathrm{ref}^{\prime}}},
\label{dmu_S}
\end{align}
By subtracting $\Delta \mu_{\mathcal{S}}$ from $\Delta \mu_{\mathrm{R}}\left(z\right)$, the following equation is obtained. 
\begin{align}
\Delta\mu_{\mathrm{R}}\left(z\right) & =\Delta\mu_{\mathcal{S}}-\dfrac{1}{\beta}\log\dfrac{\braket{\Phi_z\left(z_{\mathrm{U}},\bm{\Lambda}_{\mathrm{U}}\right)}_{\mathrm{sol}^{\prime}}}{\braket{\theta_{\mathcal{S}}\left(z_{\mathrm{U}},\bm{\Lambda}_{\mathrm{U}}\right)}_{\mathrm{sol}^{\prime}}} \notag \\
 & \quad +\dfrac{1}{\beta}\log\dfrac{\braket{\Phi_{z}\left(z_{\mathrm{U}},\bm{\Lambda}_{\mathrm{U}}\right)}_{\mathrm{ref}^{\prime}}}{\braket{\theta_{\mathcal{S}}\left(z_{\mathrm{U}},\bm{\Lambda}_{\mathrm{U}}\right)}_{\mathrm{ref}^{\prime}}}.
\end{align}
The arguments of the logarithm for the second and third terms respectively stand 
for the population ratios between region of $z$ to $z+\Delta z$ in phase R and state $\mathcal{S}$ for the solution system and for the reference solvent system. Furthermore, if we describe the permeation process using only $z$ and drop the $\bm{\Lambda}$-dependence in the above equation, one can obtain
\begin{align}
\Delta\mu_{\mathrm{R}}\left(z\right) & =\Delta\mu_{\mathcal{S}}
+\Delta G_{\mathcal{S}\to z}\left(z\right) + \dfrac{1}{\beta}\log \dfrac{\Delta z}{l_{\mathcal{S}}},
\label{eq:delta_mu_profile}
\end{align}
where 
\begin{align}
\Delta G_{\mathcal{S}\to z}\left(z\right) = -\dfrac{1}{\beta}\log\dfrac{\braket{\theta_{z}\left(z_{\mathrm{U}}\right)}_{\mathrm{sol}^{\prime}}}{\braket{\theta_{\mathcal{S}}\left(z_{\mathrm{U}}\right)}_{\mathrm{sol}^{\prime}}},
\label{dGsz}
\end{align}
and $l_{\mathcal{S}}$ is the width of state $\mathcal{S}$ along the $z$-direction. 
Thus, once we calculate $\Delta \mu_{\mathcal{S}}$, 
the profile of $\Delta \mu_{\mathrm{R}}\left(z\right)$ can be evaluated without the additional free energy calculations.
From \Eqs{Delta_G}{eq:delta_mu_profile}, $\Delta G\left(z\right)$ can be expressed as 
\begin{align}
 \Delta G\left(z\right) & =\Delta\Delta\mu
+\Delta G_{\mathcal{S}\to z}\left(z\right)
+\dfrac{1}{\beta}\log\dfrac{\Delta z}{l_{\mathcal{S}}}, \label{dGz_dmu_pmf}
\end{align}
where
\begin{align}
 \Delta \Delta \mu &= \Delta \mu_{\mathcal{S}} - \Delta \mu_{\mathrm{D}}.
 \label{ddmu}
\end{align} 
%

\section{Computational methods\label{sec:method}}
%
\subsection{System setups}
We investigated two different membrane permeation systems composed of 1-palmitoyl-2-oleoyl-\textit{sn}-glycero-3-phosphocholin (POPC) lipid bilayer and small permeant species, ethanol and methylamine.
Both the infinitely dilute and finite (1 mol\%) concentrations of the permeants were examined.
The solvent was water and the system temperature was 298.15 K.
The force fields for POPC, permeants, and water were CHARMM36,\cite{klauda2010update} CHARMM generalized force field (CGenFF),\cite{vanommeslaeghe2010charmm} and CHARMM-compatible TIP3P model,\cite{jorgensen1983comparison} respectively. 
The initial configurations of the membrane were prepared with CHARMM-GUI server.\cite{lee2018charmm,jo2008charmm,wu2014charmm}
For all the systems involving the bilayer described below, the numbers of POPC and  water were 100 per leaflet and 18000, respectively. 
The number of the permeants was unity for the dilute systems and 182 for the 1 mol\% concentration systems, respectively.
The initial configurations for water and permeants 
were prepared with Packmol.\cite{martinez2009packmol}

To compute the quantities required for the RP theory, 
we performed the MD simulations of the membrane systems in which one of the permeant molecules 
was initially located at the center of the membrane.
This permeant molecule was referred to as the tagged permeant.  
%
%
As for the computation of $\Delta G\left(z\right)$ (\Eq{Delta_G}), the BAR method was employed. 
In the case of the 1 mol\% concentration systems,  
we also performed the MD simulations of the membrane systems 
in which all the permeants were randomly placed in the solution phase 
for computing the permeability coefficient by means of the transition-based counting (TBC) method.\cite{ghysels2019permeability,venable2019molecular,davoudi2021sampling}
%
%
The equilibration scheme for the systems involving the membrane was described in \Sec{equil_memb}.  
%

%
%
%
 
All the MD simulations were performed using GENESIS 2.0.\cite{jung2015genesis,kobayashi2017genesis,jung2021new}
We employed Bussi thermostat for the temperature control in NVT/NPT ensembles 
and Bussi barostat for the pressure control in NPT ensemble.\cite{bussi2007canonical}
We employed the NVT ensemble only for the early stages of the equilibration (Table S1 of the supplementary material and \Sec{md_er}). 
The velocity Verlet integrator (VVER)\cite{swope1982computer} and reversible reference system propagator algorithm (r-RESPA)\cite{tuckerman1992reversible} were utilized.
In the case of VVER integrator, the time interval was 2 fs except for the early stages of the equilibration for the membrane systems (Table S1 of the supplementary material). 
The time interval was 2.5 fs for r-RESPA integrator.   
The Lennard-Jones (LJ) interaction was truncated by applying the switching function, with the switching range of 10--12 \AA{}.
We employed the smooth particle mesh Ewald (SPME) method for computing the electrostatic interactions.\cite{darden1993particle, essmann1995smooth}
The number of grids for the SPME method was automatically determined in GENESIS so that the grid spacing was shorter than 1.2 \AA{}.
All bonds involving hydrogen atoms were constrained by means of the SHAKE/RATTLE algorithms,\cite{ryckaert1977numerical,andersen1983rattle}
and water molecules were kept rigid using SETTLE algorithm.\cite{miyamoto1992settle} 
\subsection{Equilibration of membrane systems\label{sec:equil_memb}}
%
The scheme of equilibration for the membrane systems is described in this subsection. 
%
%
We equilibrated the systems with the NVT/NPT MD simulations (1.875 ns in total) with VVER integrator according to the GENESIS input files created in CHARMM-GUI server.
In this scheme, the $z$-coordinates of all the phosphorus atoms in the membrane were restrained using the harmonic potential with respect to their initial positions.
The harmonic potentials for the inversion angle of the glycerol group and the dihedral angle involving a double bond in acyl chains (Sec. S1 and Fig. S1 of the supplementary material) were imposed to keep the 
stereoisomeric structure and \textit{cis} form, respectively.
The force constants of these potentials were gradually decreased during the equilibration.
The detail of the equilibration is found in Table S1 of the supplementary material.
As for the simulations used for computing $P_{\mathrm{RET}}\left(t\right)$ and $k_{\mathrm{R\to A}}$, 
the harmonic potential was imposed on the $z$-component of the CoM for the tagged permeant
to locate it around the membrane center ($z=0$). 
Only the heavy atoms were considered in the calculation of the CoM, 
and the force constant for the harmonic potential was set to 1 $\mathrm{kcal~mol^{-1}~\AA^{-2}}$. 
\subsection{MD simulations for computing $P_{\mathrm{RET}}\left(t\right)$ and $k_{\mathrm{R\to A}}$\label{sec:md_ret_ra}}
We conducted the MD simulations with r-RESPA integrator for the membrane systems with the tagged permeant 
to compute $P_{\mathrm{RET}}\left(t\right)$ and $k_{\mathrm{R\to A}}$.
See \Sec{comput_thermo_and_kine} for the schemes to determine $P_{\mathrm{RET}}\left(t\right)$ and $k_{\mathrm{R\to A}}$.
From the final snapshots obtained from the equilibration (\Sec{equil_memb}),
350 ns MD (r-RESPA) simulation was performed, while imposing the following flat-bottom (FB) 
potential on the $z$-component of the CoM for the tagged permeant ($z_{\mathrm{U}}$).
\begin{align}
  U_{\mathrm{FB}}(z_{\mathrm{U}}) =
    \begin{cases}
        k(z_{\mathrm{U}}-z_1)^2, & z_{\mathrm{U}} \leq z_1, \\
        0,          & z_1 < z_{\mathrm{U}} \leq z_2,\\
        k(z_{\mathrm{U}}-z_2)^2, & z_2 < z_{\mathrm{U}}.
    \end{cases}
    \label{FB}
\end{align}
Here, $k, z_1, z_2$ were set to $10\,\mathrm{kcal~mol^{-1}~\AA^{-2}}$, 0 \AA, and 7 \AA, respectively. 
Then, we computed 300 different MD (50 ns) simulations with the FB potential (\Eq{FB}), where the random seeds for the thermostat and barostat were different among the different runs.
The final snapshots were used for the simulations to compute $P_{\mathrm{RET}}\left(t\right)$ 
and $k_{\mathrm{R\to A}}$ described below. 
For the calculation of $P_{\mathrm{RET}}\left(t\right)$, we conducted 20 ns MD simulations with the half flat-bottom (HFB) potential defined as
\begin{align}
U_{\mathrm{HFB}}^{\mathrm{RET}}\left(z_{\mathrm{U}}\right) & =\begin{cases}
k\left(z_{\mathrm{U}}-z_{1}\right)^{2}, & z_{\mathrm{U}}\leq z_{1},\\
0, & z_{1}<z_{\mathrm{U}},
\end{cases} 
\end{align}
where $k$ and $z_{1}$ were set to $10~\mathrm{kcal~mol^{-1}~\AA^{-2}}$ and $0~\mathrm{\AA}$, respectively. In the case of $k_{\mathrm{R\to A}}$, 50 and 30 ns MD simulations were performed for the ethanol and methylamine systems, respectively, 
in the presence of the following HFB potential.
\begin{align}
U_{\mathrm{HFB}}^{\mathrm{R\to A}}\left(z_{\mathrm{U}}\right) & =\begin{cases}
0, & z_{\mathrm{U}}\leq z_{2},\\
k\left(z_{\mathrm{U}}-z_{2}\right)^{2}, & z_{2}<z_{\mathrm{U}}.
\end{cases} 
\end{align}
Here, $k$ and $z_{2}$ were $10~\mathrm{kcal~mol^{-1}~\AA^{-2}}$ and $7~\mathrm{\AA}$, respectively. 
\subsection{MD simulations for computing $\Delta G\left(z\right)$\label{sec:md_er}}
%
%
We performed the BAR method with Hamiltonian replica-exchange MD (BAR/H-REMD) simulations\cite{jiang2010free} 
implemented in GENESIS\cite{oshima2022modified, matsunaga2022use} 
to compute $\Delta \Delta \mu = \Delta \mu_{\mathcal{S}} - \Delta \mu_{\mathrm{D}}$ (\Eq{ddmu}).
The integrator for the BAR/H-REMD simulations was r-RESPA.
Five configurations were randomly sampled from 300 configurations prepared for the MD simulations 
for $P_{\mathrm{RET}}\left(t\right)$ and $k_{\mathrm{R\to A}}$ described in \Sec{md_ret_ra}.
As well as in \Sec{md_ret_ra}, one permeant molecule located near the membrane center 
was treated as the tagged permeant.
The definition of state $\mathcal{S}$ was set to $0\leq z_{\mathrm{U}}/\mathrm{\AA} < 3$.
After 5 ns MD (r-RESPA) simulation for each run while imposing the FB potential (\Eq{FB}) with $k=10~\mathrm{kcal~mol^{-1}}$, $z_{1}=0~\mathrm{\AA}$, and $z_{2}=3~\mathrm{\AA}$, we conducted 
the 5 ns BAR/H-REMD simulation with 24 replicas.
During the BAR/H-REMD simulation, the same FB potential was also 
imposed on the tagged permeant for all the replicas. 
The potential energy function used in the BAR/H-REMD simulations and the scheme of computing $\Delta \Delta \mu$ 
were described in Sec. S2 and Fig. S2 of the supplementary material.
As for state D, we adopted the different schemes for the dilute and 1 mol\% concentration systems.
In the case of the dilute systems, the aqueous solutions containing a permeant molecule were prepared. 
The number of water molecules was set to 5000.
We performed the 1 ns MD (VVER, NPT) simulation for equilibration, 
followed by the 5 ns BAR/H-REMD simulation.  
Regarding the 1 mol\% concentration systems, we used the five initial configurations that were the same as those used for state $\mathcal{S}$, and  
one of the permeants existing at $45 \leq z/\mathrm{\AA} < 55$ was treated as the tagged permeant for each initial configuration.
Then, we performed the 5 ns BAR/H-REMD simulation while imposing the FB potential (\Eq{FB}) 
with $k=10~\mathrm{kcal~mol^{-1}}$, $z_{1}=45~\mathrm{\AA}$, and $z_{2}=55~\mathrm{\AA}$.

In order to calculate $\Delta G_{\mathcal{S}\to z}\left(z\right)$ (\Eq{dGsz}),
50 and 25 ns MD (r-RESPA) simulations were conducted for the ethanol and methylamine systems while imposing the FB potential with $k=10~\mathrm{kcal~mol^{-1}}$, $z_{1}=0~\mathrm{\AA}$, and $z_{2} = 7~\mathrm{\AA}$.
The number of runs was 5 for both the systems and the initial configurations were the same as those used in the BAR/H-REMD simulations.
%

%
\subsection{MD simulations for transition-based counting (TBC)\label{sec:MD_TB}}
After the equilibration of the membrane systems that contain 1 mol\% permeants (\Sec{equil_memb}), we performed the 350 ns MD simulations without any restraints.
From the final snapshot, we conducted 10 MD (50 ns for each) simulations for further equilibration. 
These runs were made distinct by assigning the different random seeds for the thermostat and barostat.
Then, we performed 200 ns MD simulation for each run as production.
r-RESPA integrator was used for the simulations described in this subsection. 
\subsection{Computation of thermodynamic and kinetic quantities for RP theory\label{sec:comput_thermo_and_kine}}
%
%
%
In the present study, we assumed that the permeation process could be described only using $z_{\mathrm{U}}$ without introducing other degrees of freedom $\bm{\Lambda}$.
%
Then, the range of $z_{\mathrm{U}}$ for state $\mathcal{S}$ was set to $0 \leq z_{\mathrm{U}}/\mathrm{\AA} \leq 3$.
From the BAR/H-REMD simulations for state $\mathcal{S}$ and phase D, 
we computed the difference of the solvation free energy, $\Delta \Delta \mu$ (\Eq{ddmu}). 
The configurations in the BAR/H-REMD simulations for state $\mathcal{S}$ 
that satisfy $ 0 \leq z_{\mathrm{U}}/\mathrm{\AA} < 3$ were used for computing $\Delta \Delta \mu$. 
%
%
%

%
To calculate the profile of $\Delta G\left(z\right)$ from \Eq{dGz_dmu_pmf}, 
we computed $\Delta G_{\mathcal{S}\to z}\left(z\right)$ (\Eq{dGsz}) 
from the trajectories of the solution system composed of the membrane, one permeant molecule, and water molecules with 181 permeant molecules for the 1 mol\% systems. 
The standard error of $\Delta G_{\mathcal{S}\to z}\left(z\right)$ was estimated using the Monte-Carlo (MC) bootstrap method.\cite{efron1992bootstrap}
The number of bootstrap samples generated by selecting the trajectories was 1000.
By using the trajectories of the MD simulations with $U_{\mathrm{HFB}}^{\mathrm{RET}}\left(z_{\mathrm{U}}\right)$ and with $U_{\mathrm{HFB}}^{\mathrm{R\to A}}\left(z_{\mathrm{U}}\right)$ described in \Sec{md_ret_ra}, we computed $P_{\mathrm{RET}}\left(t\right)$ and $k_{\mathrm{R\to A}}$, respectively. 
Let us define the time series of the characteristic function for phase R for the $\alpha$th trajectory as $\Theta^{\left(\alpha\right)}\left(t\right)$.
Then, we discretize time $t$ as $t_{k}= k\Delta t~(k=0,1,\cdots, N_{\mathrm{step}}-1)$, where $\Delta t$ is the time interval.
In the present study, $\Delta t$ was set to 0.25 ps.
$N_{\mathrm{step}}$ is the number of time steps in a trajectory. 
$P_{\mathrm{RET}}\left(t_{k}\right)$ was computed with the following equation.
\begin{align}
& P_{\mathrm{RET}}\left(t_{k}\right)  \notag \\
& =\dfrac{N_{\mathrm{step}}}{N_{\mathrm{step}}-k}
\dfrac{\displaystyle \sum_{\alpha=1}^{N_{\mathrm{traj}}}\sum_{l=0}^{N_{\mathrm{step}}-k-1}\Theta^{\left(\alpha\right)}\left(t_{k}+t_{l}\right)\Theta^{\left(\alpha\right)}\left(t_{l}\right)}
{\displaystyle\sum_{\alpha=1}^{N_{\mathrm{traj}}}\sum_{l=0}^{N_{\mathrm{step}}-1}\Theta^{\left(\alpha\right)}\left(t_{l}\right)}. 
\end{align}
We computed $k_{\mathrm{R\to A}}$ based on the frequency of the transition from $\mathrm{R}$ to $\mathrm{A}$ as
\begin{align}
k_{\mathrm{R\to A}} & =\dfrac{\displaystyle\sum_{\alpha=1}^{N_{\mathrm{traj}}}\delta_{\mathrm{R\to A}}^{\left(\alpha\right)}}{\displaystyle \sum_{\alpha=1}^{N_{\mathrm{traj}}}\sum_{l=0}^{N_{\mathrm{step}}^{\left(\alpha\right)}-1}\Theta^{\left(\alpha\right)}\left(t_{l}\right)\Delta t},
\end{align}
where $\delta_{\mathrm{R\to A}}^{\left(\alpha\right)}$ is a characteristic function for transition, which is unity when the transition event is observed in the $\alpha$th trajectory and vanishes otherwise.
$N_{\mathrm{step}}^{\left(\alpha\right)}$ is the number of time steps until the transition event is observed for the first time in the $\alpha$th trajectory.
The entry of the permeant into region $z \leq -25 ~\mathrm{\AA}$ was regarded as the transition. 
We employed the MC bootstrap method for the error estimations of $P_{\mathrm{RET}}\left(t\right)$ and $k_{\mathrm{R\to A}}$. 
The number of the bootstrap samples generated by selecting the trajectories was 1000.
\subsection{Transition-based counting (TBC) method}
According to the transition-based counting (TBC) method,\cite{ghysels2019permeability,venable2019molecular}
the permeability coefficient is given by 
\begin{align}
  \mathcal{P}_{\mathrm{ss}}^{\mathrm{TBC}} = \dfrac{r}{2c_{\mathrm{w}}},
  \label{permeability_counting}
\end{align}
where $r$ is the rate of the permeants for passing through the membrane per unit area and time, and $c_{\mathrm{w}}$ is the average concentration of the permeants outside the membrane, and the superscript TBC signifies transition-based counting.
For the computation of $r$ and $c_{\mathrm{w}}$, the membrane region was defined as $\left|z\right| < 20~\mathrm{\AA}$.
The definition of the membrane region was same as that used in the previous study.\cite{Ghorbani_2020} 
We performed the error analysis of $\mathcal{P}^{\mathrm{TB}}_{\mathrm{ss}}$ by means of the MC bootstrap method.
The number of the bootstrap samples generated by selecting the trajectories was 1000.
%
%

\section{RESULTS AND DISCUSSION}
\subsection{Free energy profiles}
\begin{figure}[t]
  \includegraphics[width=1.0\linewidth]{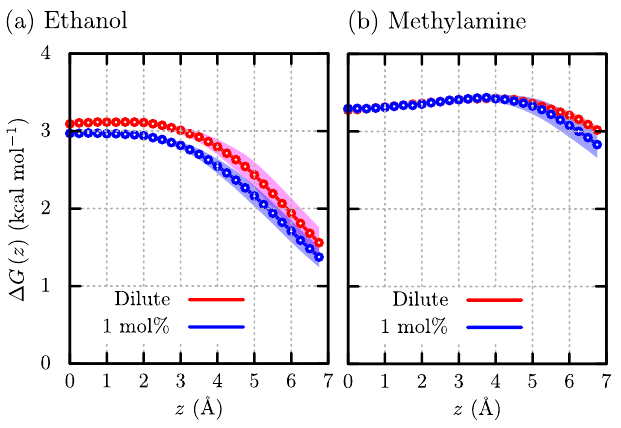}
  \caption{Free energy profile along $z$-direction, $\Delta G\left(z\right)$, near the membrane center ($z=0$) for (a) ethanol and (b) methylamine. The colored regions indicate the statistical uncertainty (standard error).\label{fig:dGz}}
\end{figure}
\begin{figure}[t]
  \centering
  \includegraphics[width=1.0\linewidth]{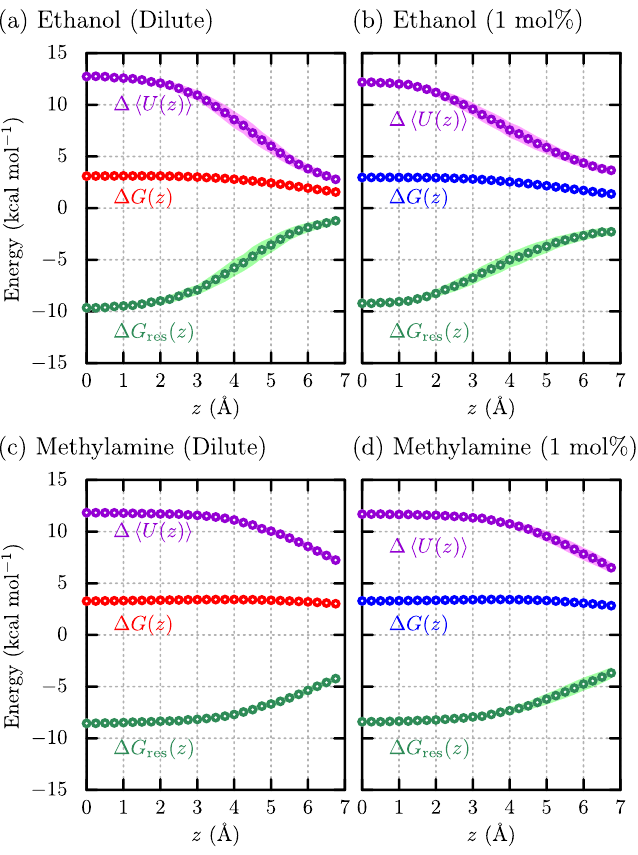}
  \caption{Profiles of $\Delta G\left(z\right)$ and its decomposition into the interaction energy part ($\Delta \braket{U\left(z\right)}$) and residual part ($\Delta G_{\mathrm{res}}\left(z\right)$) based on \Eq{dG_decomp} for (a) ethanol (dilute), (b) ethanol (1 mol\%), (c) methylamine (dilute), and (d) methylamine (1 mol\%). The colored regions indicate the statistical uncertainty (standard error).\label{fig:dG_decomp}}
\end{figure}
%
We first examine the free energy profile inside the membrane along the $z$-direction, $\Delta G\left(z\right)$, 
using \Eq{dGz_dmu_pmf} (\Fig{dGz}).
In the case of the dilute ethanol system (\Fig{dGz}(a)), 
the free energy barrier with respect to phase D (corresponding to $z=\infty$) is located at the membrane center ($z=0$).
The observed location of the barrier is typical for the hydrophilic permeants, because the inner region ($|z|\leq 10~\mathrm{\AA}$) composed of the hydrophobic acyl chains of POPC is energetically unfavorable for such a permeant.\cite{venable2019molecular, bemporad2004permeation}
The barrier height is $\sim${}$3.1~\mathrm{kcal~mol^{-1}}$.
In the presence of 1 mol\% ethanol, it is found that the height is slightly lowered to $\sim${}$3.0~\mathrm{kcal~mol^{-1}}$.
The PMF, $\Delta G_{\mathrm{PMF}}\left(z\right)$, obtained using the MD simulations for transition-based counting (\Sec{MD_TB}) gives a similar height 
for the 1 mol\% ethanol system (Fig. S3(a) of the supplementary material).  
A previous MD study also reported a decrease in the height with increasing ethanol concentration in the range from 1 mol\% to 18 mol\%.\cite{Ghorbani_2020} 
Regarding the dilute methylamine system (\Fig{dGz}(b)), $\Delta G\left(z\right)$ exhibits a shallow minimum around $z=0$ with the height of $\sim${}$3.3~\mathrm{kcal~mol^{-1}}$ and the barrier located at $z\sim${}$4.5~\mathrm{\AA}$. 
The similar behavior was also reported by Bemporad \textit{et al.}\cite{bemporad2004permeation} 
The barrier height is found to be hardly changed even for the 1 mol\% methylamine system.
As well as in the case of the 1 mol\% ethanol system, $\Delta G\left(z\right)$ 
is almost the same as that from $\Delta G_{\mathrm{PMF}}\left(z\right)$ within the range of $0\leq z/\mathrm{\AA} \leq 7$ (Fig. S3(b) of the supplementary material), indicating the validity of the scheme of computing the free energy profile through \Eq{dGz_dmu_pmf}.

For further analysis, we decompose $\Delta G\left(z\right)$. 
According to the classical density functional theory,\cite{Matubayasi_2019} the solvation free energies can be exactly decomposed as follows.
\begin{align}
\Delta\mu_{\mathrm{D}} & =\braket{U_{\mathrm{D}}}+\Delta\mu_{\mathrm{D,res}},\\
\Delta\mu_{\mathrm{R}}\left(z\right) & =\braket{U_{\mathrm{R}}\left(z\right)}+\Delta\mu_{\mathrm{R,res}}\left(z\right).
\end{align}
Here, $U_{i}$  and $\Delta \mu_{i,\mathrm{res}}$ ($i=\mathrm{D~or~R}$) are respectively the interaction energy of a permeant with the surrounding environment and the residual part of $\Delta \mu_{i}$ which is composed of the pair entropy and many-body terms.
$\Delta\mu_{i,\mathrm{res}}$ corresponds to the free energy penalty due to the structural changes of the surrounding environment upon solvation.
$\Delta G\left(z\right)$ can be decomposed into the contributions from the interaction energy and residual part as
\begin{align}
 \Delta G\left(z\right) & =\Delta\braket{U\left(z\right)}+\Delta G_{\mathrm{res}}\left(z\right),
 \label{dG_decomp}
\end{align}
where $\Delta \braket{U\left(z\right)} = \braket{U_{\mathrm{R}}\left(z\right)} - \braket{U_{\mathrm{D}}}$ and $\Delta G_{\mathrm{res}}\left(z\right) = \Delta \mu_{\mathrm{R,res}}\left(z\right) - \Delta \mu_{\mathrm{D,res}}$.
The profiles of $\Delta G\left(z\right)$, $\Delta\braket{U\left(z\right)}$, and $\Delta G_{\mathrm{res}}\left(z\right)$ are shown in \Fig{dG_decomp}.
For both the ethanol and methylamine systems, $\Delta \braket{U\left(z\right)}$ and $\Delta G_{\mathrm{res}}\left(z\right)$ respectively demonstrate positive and negative contributions to $\Delta G\left(z\right)$ within the inner region, regardless of the permeant concentration. 
Given the hydrophilic nature of these permeants, the positive value of $\Delta \braket{U\left(z\right)}$ evidently 
arises from the dehydration penalty. 
The negative value of $\Delta G_{\mathrm{res}}\left(z\right)$ inside the membrane 
indicates that the free-energy penalty brought by the structural change of the membrane is smaller 
than that of the solvent water at phase D.
As shown by Cardenas and Elber\cite{cardenas2014modeling} as well as by Chipot and Comer\cite{chipot2016subdiffusion} through the MD simulations, the voids in the inner region are highly populated compared with the bulk solvent.
The presence of voids in the inner region could mitigate the structural change of the membrane upon solvation of the permeant, leading to a decrease in $\Delta G_{\mathrm{res}}\left(z\right)$.   
For both the concentrations, the value of $\Delta \braket{U\left(z\right)}$ at $z=0$ for ethanol is larger than for methylamine, reflecting the higher hydrophilicity of ethanol. 
The smaller value of $\Delta G_{\mathrm{res}}\left(z\right)$ at $z=0$ for ethanol compared to methylamine may be attributed to ethanol's larger molecular size.
As the permeant concentration increases, 
$\Delta \braket{U\left(z\right)}$ and $\Delta G_{\mathrm{res}}\left(z\right)$ at $z=0$ respectively decrease and increase for ethanol, while there is little change in these quantities for methylamine. 
\subsection{Kinetics of returning and crossing processes}
\begin{figure}[t]
  \centering
  \includegraphics[width=1.0\linewidth]{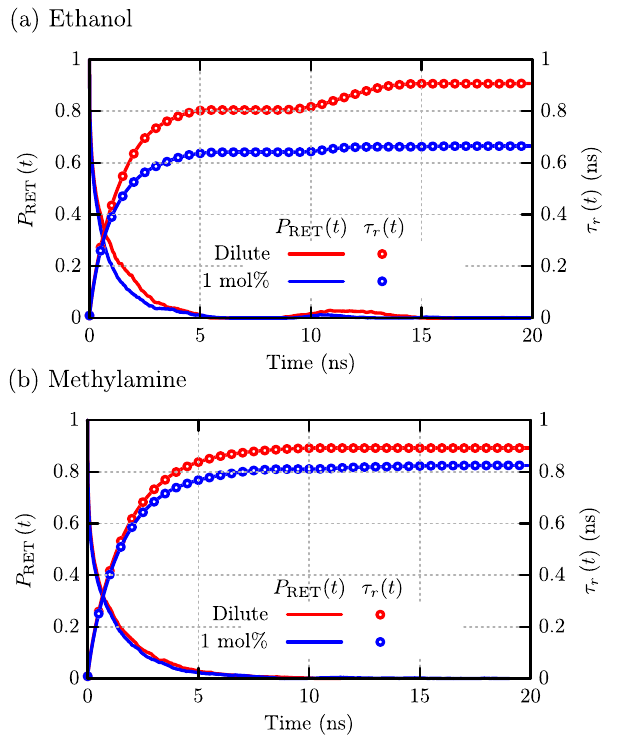}
  \caption{Returning probability $P_\mathrm{RET}(t)$, and its running integral $\tau_r(t)$ for (a) ethanol and (b) methylamine systems. The $z$-range of phase R is set to $3~\mathrm{\AA}$. 
  \label{fig:Pret}}
\end{figure}
\begin{figure}[t]
  \centering
  \includegraphics[width=1.0\linewidth]{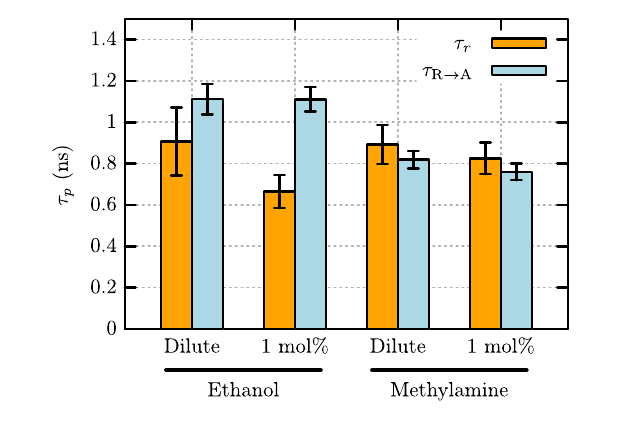}
  \caption{Time constant of the dissociation, $\tau_{r}$, and that of  
  the transition from phase R to A (crossing), $\tau_{\mathrm{R\to A}}$.
  $\tau_{r}$ is defined as $\tau_{r} = \tau_{r}\left(\infty\right)$. 
  The error bars are provided at the standard error. 
  \label{fig:tau}
  }
\end{figure}
In this subsection, we discuss the kinetic property of the permeants at phase R.
We define the $z$-range of phase R ($\bm{\Upsilon}$) as
\begin{align}
 \bm{\Upsilon} = \left\{z~|~0\leq z/\mathrm{\AA}\leq 3\right\}, 
\end{align}
for both the ethanol and methylamine systems.
The lower bound is fixed to $z=0$, corresponding to the membrane center, and upper bound is decided so that the permeability coefficients are hardly changed by the variation in the upper bound, as will be discussed in \Sec{Pcoeff}. 

The returning probability $P_\mathrm{RET}(t)$ and its running integral defined as 
\begin{align}
  \tau_{r}\left(t\right) = \int_{0}^{t}ds\, P_{\mathrm{RET}}\left(s\right),
  \label{running_integral}
\end{align}
are plotted in \Fig{Pret}.
%
%
Note that $\tau_{r}\left(\infty\right)$ is the dissociation time constant from phase R.\cite{Kasahara_2021}
For the dilute ethanol system, $P_{\mathrm{RET}}\left(t\right)$ is observed to converge 
to zero at $\sim${}$5$ ns, followed by a subsequent redistribution occurring at $\sim${}$11$ ns.
$\tau_{r}\left(t\right)$ is converged after $\sim${}$15$ ns.
The first decay of $P_{\mathrm{RET}}\left(t\right)$ becomes faster for the 1 mol\% ethanol system.
In addition, the redistributive behavior observed for the dilute system is weakened, 
resulting in the faster convergence of $\tau_{r}\left(t\right)$. 
It is well-known that a stable state for ethanol exists between the membrane center and phase D, \cite{comer2017permeability}
and the stability of this state is suppressed with increasing ethanol concentration.\cite{Ghorbani_2020}
Thus, the return of ethanol to phase R could be facilitated by the trap of ethanol at the stable state within the membrane, emphasized under conditions where the stability of this state is high.
In the cases of the dilute and 1 mol\% methylamine systems, 
it is seen that $P_{\mathrm{RET}}\left(t\right)$ converges to zero at $\sim${}$7$ ns. 
Unlike ethanol, methylamine does not show the redistributive behavior. 
As shown in the profile of $\Delta G_{\mathrm{PMF}}\left(z\right)$ for the 1 mol\% system (Fig. S3(b) of the supplementary material), the stable state is present inside the membrane, but the stability is lower than that for the ethanol.
This can be reason why the redistributive behavior is not observed for the methylamine systems. 

The time constant of the dissociation, $\tau_{r}$, and that of the transition from phase R to A (crossing), $\tau_{\mathrm{R\to A}}$,
are shown in \Fig{tau}.
Here, we define $\tau_{r}$ and $\tau_{\mathrm{R\to A}}$ as $\tau_{r} = \tau_{r}\left(\infty\right)$ 
and $\tau_{\mathrm{R\to A}} = 1 / k_{\mathrm{R\to A}}$, respectively. 
For the dilute ethanol system, $\tau_{r}$ is smaller than $\tau_{\mathrm{R\to A}}$, 
indicating that dissociation from phase R preferentially occurs over crossing. 
This trend reflects the downhill profile of $\Delta G\left(z\right)$ around $z=0$ (\Fig{dGz}(a)).
In the presence of 1 mol\% ethanol,
a reduction in $\tau_{r}$ is discernible as also shown in \Fig{Pret}, while $\tau_{\mathrm{R\to A}}$ remains largely unchanged from that in the dilute system.
As for the methylamine systems, the difference between $\tau_{r}$ and $\tau_{\mathrm{R\to A}}$ is negligibly small under the dilute condition.
Furthermore, both $\tau_{r}$ and $\tau_{\mathrm{R\to A}}$ hardly change with increasing concentration, 
consistent with the fact that the profile of $\Delta G\left(z\right)$ is not dependent on the concentration (\Fig{dGz}(b)).  

\subsection{Permeability coefficient\label{sec:Pcoeff}}
\begin{figure}[t]
  \centering
  \includegraphics[width=1.0\linewidth]{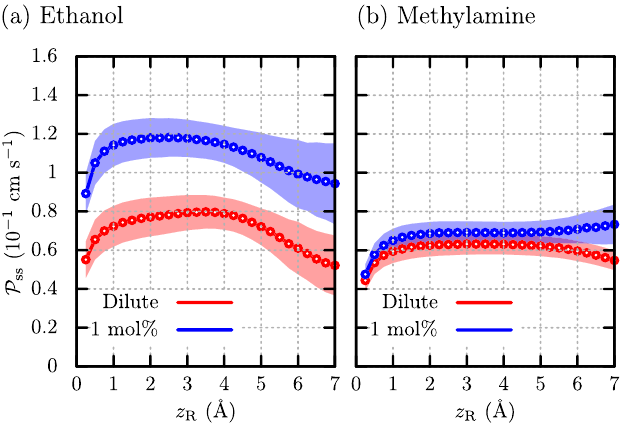}
  \caption{Dependency of the permeability coefficients ($\mathcal{P}_{\mathrm{ss}}$) obtained from the RP theory (\Eq{Pss_final}) on the choice of state R ($\bm{\Upsilon}=\left\{z~|~0\leq z \leq z_{\mathrm{R}}\right\}$) for (a) ethanol and (b) methylamine. The colored regions indicate the statistical uncertainty (standard error).}
  \label{fig:Permeability}
\end{figure}
%
%
\begin{table*}[t]
\renewcommand{\arraystretch}{1.5}
\centering
\caption{Permeability coefficients ($\mathcal{P}_{\mathrm{ss}}$) computed from the RP theory and from the transition-based counting (TBC) method. The thermodynamic ($K^*$) and kinetic ($\chi$) contributions are also shown. The errors are provided at the standard error, and the errors for $K^*$ are not shown since they are smaller than $0.001~\mathrm{\AA}$.\label{tab:Pss}}
\begin{tabular*}{13cm}{@{\extracolsep{\fill}}ccccccc}
\hline
\hline 
 &  & \multicolumn{2}{c}{$\mathcal{P}_{\mathrm{ss}}\,\left(\mathrm{cm\,s^{-1}}\right)$} &  &  & \tabularnewline
\cline{3-4} \cline{4-4} 
 &  & RP & TBC &  & $K^{*}\,\left(\mathrm{\text{\AA}}\right)$ & $\chi\,\left(\mathrm{ns}^{-1}\right)$\tabularnewline
\hline 
Ethanol & Dilute & $0.079\pm0.009$ & $\cdots$ &  & $0.016$ & $0.50\pm0.05$\tabularnewline
 & 1 mol\% & $0.12\pm0.01$ & $0.18\pm0.03$ &  & $0.021$ & $0.56\pm0.04$\tabularnewline
\hline 
Methylamine & Dilute & $0.063\pm0.005$ & $\cdots$ &  & $0.011$ & $0.58\pm0.05$\tabularnewline
 & 1 mol\% & $0.069\pm0.006$ & $0.052\pm0.005$ &  & $0.011$ & $0.64\pm0.05$\tabularnewline
\hline
\hline 
\end{tabular*}
\end{table*}
In order to calculate the permeability coefficients at steady state, $\mathcal{P}_{\mathrm{ss}}$, 
through the RP theory (\Eq{Pss_final}),
the definition of phase R ($\bm{\Upsilon}$) is required. 
Then, we investigate the dependency of $\mathcal{P}_{\mathrm{ss}}$
on the choice of the $z$-range of $\bm{\Upsilon}$.
Let us express $\bm{\Upsilon}$ as follows.
\begin{align}
 \bm{\Upsilon} = \left\{z~|~0 \leq z \leq z_{\mathrm{R}}\right\}. 
\end{align} 
\Fig{Permeability} plots $\mathcal{P}_{\mathrm{ss}}$ as a function of $z_{\mathrm{R}}$. 
In the case of the ethanol systems, it is observed that $\mathcal{P}_{\mathrm{ss}}$ 
exhibits a slight dependency on $z_{\mathrm{R}}$ 
when $z_{\mathrm{R}} < 1.5~\mathrm{\AA}$ and $z_{\mathrm{R}} > 3.5~\mathrm{\AA}$ both under the dilute and 1 mol\% conditions.
$\mathcal{P}_{\mathrm{ss}}$ for the methylamine systems remains constant at 
$1.5~\mathrm{\AA} < z_{\mathrm{R}} < 5~\mathrm{\AA}$ 
irrespective of the concentration. 
Although the theoretical analysis using the Smoluchowski equation claims 
that the RP theory becomes accurate when $\bm{\Upsilon}$ is sufficiently narrow,\cite{kim2009rigorous}
the $z_{\mathrm{R}}$-dependency of $\mathcal{P}_{\mathrm{ss}}$ is discernible at too small $z_{\mathrm{R}}$ values 
(e.g., $z_{\mathrm{R}} = 0.5~\mathrm{\AA}$).
This observation suggests that the local equilibrium within $\bm{\Upsilon}$ assumed in the RP theory 
might be violated for such a condition (Appendix A). 
The contraction of the multiple state-to-state transitions involved in the crossing process ($\mathrm{R}\to \mathrm{A}$) 
to a two-state transition, as represented by $k_{\mathrm{R\to A}}$, appears to cause 
the $z_{\mathrm{R}}$-dependency. 
However, from the analysis of the permeability corresponding to the arrival of a permeant at the metastable state  
within the membrane ($z${}$\sim${}$-15~\mathrm{\AA}$) from phase D, 
we confirm that the presence of the multiple states 
is not relevant with the $z_{\mathrm{R}}$-dependency of $\mathcal{P}_{\mathrm{ss}}$ 
(Sec. S3 and Fig. S4 of the supplementary material).  
In a previous study on protein-ligand binding with the RP theory,\cite{kasahara2023elucidating} 
the appropriate $\bm{\Upsilon}$ was determined so that the binding rate constants are hardly affected by the variation 
in $\bm{\Upsilon}$.
Similarly, we set $z_{\mathrm{R}}$ to $3~\mathrm{\AA}$, at which the profiles of $\mathcal{P}_{\mathrm{ss}}$ along $z_{\mathrm{R}}$ are almost flat for both the ethanol and methylamine systems. 
%
%

%
%
%
The values of $\mathcal{P}_{\mathrm{ss}}$ using the aforementioned definition of $\bm{\Upsilon}$ are summarized in \Table{Pss}, 
together with those obtained from the transition-based counting (TBC) method for the 1 mol\% systems.
For the 1 mol\% systems, 
%
the RP theory predicts a larger $\mathcal{P}_{\mathrm{ss}}$ for ethanol ($0.12\pm 0.01~\mathrm{cm~s^{-1}}$) than for methylamine ($0.069\pm 0.006~\mathrm{cm~s^{-1}}$), consistent with the TBC method.
Furthermore, the values for $\mathcal{P}_{\mathrm{ss}}$ are satisfactorily close to those obtained from the TBC method [$0.18\pm 0.03~\mathrm{cm~s^{-1}}$ and $0.052\pm 0.005~\mathrm{cm~s^{-1}}$ for ethanol and methylamine, respectively], revealing the validity of the RP theory.
To realize the further agreement of $\mathcal{P}_{\mathrm{ss}}$ 
between the RP and the TBC,
employing an improved expression of $\mathcal{P}_{\mathrm{ss}}$ 
which accounts for the non-Markovianity partly, derived through the perturbative expansion, 
could be a promising approach (\Eq{Improved_Pss} in \Appendix{Improved_Pss}).\cite{kim2009rigorous}
The difference in $\mathcal{P}_{\mathrm{ss}}$ between the RP and the TBC, observed for the 1 mol\% ethanol system, 
may arise from the inadequacy of describing 
the underlying dynamics in phase R solely with the $z$-coordinate.
Using such an inadequate coordinate as the reaction coordinate 
brings the non-Markovianity of the dynamics. 
Thus, the improved expression of $\mathcal{P}_{\mathrm{ss}}$ (\Eq{Improved_Pss}) could overcome this challenge. 
The improved expression  involves a multiple-time correlation function, 
indicating the necessity of developing an efficient computational method 
for evaluating such a function when utilizing this expression. 

We examine the concentration dependence of $\mathcal{P}_{\mathrm{ss}}$. 
An increase in $\mathcal{P}_{\mathrm{ss}}$ is observed for the ethanol systems as ethanol concentration increases, 
while such a change is hardly discernible for the methylamine systems.
This trend is consistent with the experimental observation that ethanol enhances the membrane permeability of drug compounds.
\cite{pershing1990mechanism}
It is well known that ethanol strongly affects the properties of lipid membranes.
For instance, an increase in ethanol concentration leads to the expansion of the area per lipid (APL) 
and the reduction of the lipid ordering,\cite{patra2006under,gurtovenko2009interaction,konas2015biophysical} 
bringing to the enhanced membrane permeability.
We confirm that the APL becomes larger as increasing ethanol concentration, 
while the effect of the 1 mol\% methylamine on the APL is 
hardly observed (Fig. S5 of the supplementary material).
To further analyze the dependence of the permeability on the permeant concentration, 
we conduct a systematic examination based on the theoretical expression of $\mathcal{P}_{\mathrm{ss}}$ 
provided by the RP theory (\Eq{Pss_final}). By defining 
\begin{align}
 \chi = \left(\tau_{\mathrm{R\to A}} + \tau_{r}\right)^{-1}, \label{chi} 
\end{align}
\Eq{Pss_final} can be rewritten as
\begin{align}
 \mathcal{P}_{\mathrm{ss}} = \chi K^{*}.
\end{align}
The above expression indicates that $K^{*}$ and $\chi$ represent the thermodynamic and 
kinetic contributions to $\mathcal{P}_{\mathrm{ss}}$, respectively.
The computed values for these quantities are listed in \Table{Pss}.
In the case of the ethanol systems, it is found that the increase in $K^{*}$ predominantly contributes 
to enhancing ethanol permeability 
with increasing ethanol concentration. 
Regarding the methylamine systems, on the other hand,  
the dependence of both $K^{*}$ and $\chi$ on methylamine concentration is negligibly small, 
resulting in the insensitivity of $\mathcal{P}_{\mathrm{ss}}$ to changes in concentration.  
%
%


\section{Conclusion}
In this study, we proposed an MD-based methodology to 
compute the permeability coefficients at steady state, $\mathcal{P}_{\mathrm{ss}}$, 
using the returning probability (RP) theory. 
The permeability coefficients represent the efficiency of the permeation processes from the donor (D) phase to acceptor (A) phase. 
Starting from the Liouville equation with the reaction sink term describing 
the local motion toward phase A in the reactive (R) phase located inside the membrane, 
we derived a formally exact expression of the permeability coefficient at unsteady state, 
which is mathematically parallel to that of the unsteady rate coefficient for molecular binding in the RP theory. 
This parallelism enabled us to employ the perturbative-expansion technique in the RP theory to yield the tractable expression of the permeability coefficient at steady state.
The resultant expression is composed of the thermodynamic and kinetic properties of phase R that can be evaluated 
through the MD simulations.
%
%
%

The present method was applied to the permeation of ethanol and methylamine through the lipid bilayer composed of POPC.
The dilute and 1 mol\% permeant systems were examined.
The thermodynamic stability analysis 
showed that both ethanol and methylamine were destabilized around phase R (membrane center) compared with phase D (solution phase), reflecting the hydrophilic nature of these permeants.
The free energy barrier was observed to decrease with increasing ethanol concentration for ethanol, 
while this effect was barely discernible for methylamine.
Regarding the kinetic properties, the dissociation of ethanol 
from phase R was promoted in the presence of 1 mol\% ethanol.
In contrast, 
the kinetics of methylamine around phase R 
remained unchanged due to the change in methylamine concentration. 
The present method yielded a larger value of $\mathcal{P}_{\mathrm{ss}}$ 
for ethanol than for methylamine under the 1 mol\% condition, 
consistent with the prediction from the transition-based counting (TBC) method. 
Furthermore, by decomposing $\mathcal{P}_{\mathrm{ss}}$ into the thermodynamic and kinetic contributions, we clarified that a concentration dependency of $\mathcal{P}_{\mathrm{ss}}$ observed  
for the ethanol systems was attributed to the sensitivity of the free energy barrier against the concentration within the inner region of the membrane. 

Since no assumption was imposed on the condition of the solution (donor and acceptor) phases in the present method, the inhomogeneous donor and acceptor phases 
such as crowded solutions\cite{nawrocki2019clustering} can 
be treated by means of the present method.
The unstirred water layer (UWL), which is an inhomogeneous region proximal to biomembranes in gastrointestinal environments and  
exerts a significant influence on the permeation of small molecules, can be another target of investigation of the RP method.\cite{wilson1971unstirred, korjamo2009analysis, kang2024impact}
%

To extend the applicability of the present method 
to long-timescale permeation phenomena, refining the method is essential.
In the present study, 
the computation of the returning probability ($P_{\mathrm{RET}}\left(t\right)$) 
and the rate constant for the transition from phase R to A ($k_{\mathrm{R\to A}}$)
was performed using a number of short MD trajectories starting from phase R.
When the kinetics around phase R is slow,  
the required length of time for each trajectory becomes longer, 
leading to an increase in computational costs.
Recently, the methodologies of efficiently 
computing the time correlation functions for the state-to-state transitions from short MD trajectories 
have been proposed based on the generalized Langevin equation (GLE).\cite{Cao2020, dominic2023memory, kasahara2022constructing}
Incorporating such methodologies could overcome the challenge in the present method.
We believe that the present method and its extension would offer a promising route 
to unveil the detailed mechanisms of permeation phenomena 
in complex biological systems, such as cellular and intestinal membranes.  
%
%

%
\begin{acknowledgments}
This work is supported by the Grants-in-Aid for Scientific Research (Nos. JP21H05249, JP23K27313, and JP23K26617)
from the Japan Society for the Promotion of Science, 
by the Fugaku Supercomputer Project (Nos. JPMXP1020230325 and JPMXP1020230327) and the Data-Driven Material Research Project (No. JPMXP1122714694) from the Ministry of Education, Culture, Sports, Science, and Technology, and by Maruho Collaborative Project for Theoretical Pharmaceutics. 
The simulations were conducted using TSUBAME3.0 and TSUBAME4.0 at Tokyo Institute of Technology, and Fugaku at RIKEN Advanced Institute for Computational Science through the HPCI System Research Project (Project IDs: hp230101, hp230205, hp230212, hp230158, hp240223, hp240224, hp240195, and hp240111).
\end{acknowledgments}
\section*{Supplementary material}
The supplementary material contains the simulation protocols, 
the scheme of the Bennett acceptance ratio (BAR) method, 
the structure of 1-palmitoyl-2-oleoyl-\textit{sn}-glycero-3-phosphocholin (POPC), 
the set of coupling parameters used in the BAR method,    
the permeability coefficient corresponding to the arrival of a permeant at the metastable state,
the potentials of mean force (PMF) along the $z$-direction, and the distributions of area per lipid (APL).
\section*{Conflict of interest}
The authors have no conflicts to disclose.
\section*{Data Availability}
The data that support the findings of this study are available from the corresponding authors upon reasonable request.
\appendix
\section{Derivation of theoretical expression of $\mathcal{P}_{\mathrm{ss}}$\label{sec:Derive_Pss}}
In this appendix, we derive a theoretical expression of the permeability coefficient, $\mathcal{P}_{\mathrm{ss}}$, by utilizing the mathematical techniques employed in the RP theory.\cite{Kasahara_2021,kim2009rigorous}

Since the system is in equilibrium at $t=0$, $\Psi_{i}\left(\bm{\Gamma},0\right)$ is equivalent to the equilibrium phase space density, $\Psi_{\mathrm{eq}}\left(\bm{\Gamma}\right)$. 
Thus, the formal solution of \Eq{dPsi} is given by
\begin{align}
\Psi_{i}\left(\bm{\Gamma},t\right) & =e^{-\left(\mathcal{L}+S\right)t}\Psi_{\mathrm{eq}}\left(\bm{\Gamma}\right), \label{Psi_formal}
\end{align}
Using the operator identity 
\begin{align}
e^{-\left(\mathcal{L}+S\right)t}&=e^{-\mathcal{L}t}-\int_{0}^{t}d\tau\,e^{-\left(\mathcal{L}+S\right)\tau}Se^{-\mathcal{L}\left(t-\tau\right)},
\label{Mori_identity} 
\end{align}
\Eq{Psi_formal} is rewritten as
\begin{align}
\Psi_{i}\left(\bm{\Gamma},t\right) & =\Psi_{\mathrm{eq}}\left(\bm{\Gamma}\right)-\int_{0}^{t}d\tau\,e^{-\left(\mathcal{L}+S\right)\tau}S\Psi_{\mathrm{eq}}\left(\bm{\Gamma}\right),
\end{align}
where we have used the relationship given by $\mathcal{L}\Psi_{\mathrm{eq}}\left(\bm{\Gamma}\right)=0$.
The above equation enables us to decompose the nonequilibrium distribution function of the permeants, $g\left({\bf r},\bm{\Lambda},t\right)$ (\Eq{g_rl}), into the equilibrium ($g_{\mathrm{eq}}\left({\bf r},\bm{\Lambda}\right)$) and nonequilibrium ($\Delta g\left({\bf r},\bm{\Lambda},t\right)$) parts as
\begin{align}
&g\left({\bf r},\bm{\Lambda},t\right) =g_{\mathrm{eq}}\left({\bf r},\bm{\Lambda}\right)+\Delta g\left({\bf r},\bm{\Lambda},t\right), \label{gr_decomp}\\
&g_{\mathrm{eq}}\left({\bf r},\bm{\Lambda}\right) =\braket{\braket{\delta\left({\bf r}-{\bf r}_{i}\right)\delta\left(\bm{\Lambda}-\bm{\Lambda}_{i}\right)}}, \label{geq_app}\\
&\Delta g\left({\bf r},\bm{\Lambda}, t\right) \notag \\
&=-\int_{0}^{t}d\tau\,\braket{\braket{\delta\left({\bf r}-{\bf r}_{i}\right)\delta\left(\bm{\Lambda}-\bm{\Lambda}_{i}\right)e^{-\left(\mathcal{L}+S\right)\tau}S}}. \label{dg_app}
\end{align}
Here, we have introduced the following notation. 
\begin{align}
\braket{\braket{\cdots}} & =\dfrac{1}{c_{\mathrm{D}}}\sum_{i=1}^{N_{0}}\int d\bm{\Gamma}\,\left(\cdots\right)\Psi_{\mathrm{eq}}\left(\bm{\Gamma}\right).
\end{align}
Substitution of Eqs. \eqref{gr_decomp}--\eqref{dg_app} into \Eq{P_exact} leads to  
\begin{align}
\mathcal{P}\left(t\right) & =\mathcal{P}_{\mathrm{eq}}\left(1-\int_{0}^{t}d\tau\,h\left(\tau\right)\right), 
\end{align}
where $\mathcal{P}_{\mathrm{eq}}$ is the equilibrium permeability coefficient defined as 
\begin{align}
\mathcal{P}_{\mathrm{eq}} & =\dfrac{1}{\sigma}\int d{\bf r}\int d\bm{\Lambda}\,S\left(z,\bm{\Lambda}\right)g_{\mathrm{eq}}\left({\bf r},\bm{\Lambda}\right) \notag \\
 & = \dfrac{1}{\sigma}\braket{\braket{S}},\label{Peq}
\end{align}
and $h\left(t\right)$ is defined as
\begin{align}
h\left(t\right) & =\dfrac{\braket{\braket{Se^{-\left(\mathcal{L}+S\right)t}S}}}{\braket{\braket{S}}}.
\end{align}
The Laplace transform ($t\to s$) of $\mathcal{P}\left(t\right)$, $\hat{\mathcal{P}}\left(s\right)$, is represented as 
\begin{align}
s\hat{\mathcal{P}}\left(s\right) & =\mathcal{P}_{\mathrm{eq}}\left(1-\hat{h}\left(s\right)\right),
\label{Noyes}
\end{align}
where $\hat{h}\left(s\right)$ is the Laplace transform of $h\left(t\right)$ given as
\begin{align}
\hat{h}\left(s\right) & =\dfrac{\braket{\braket{S\left(s+\mathcal{L}+S\right)^{-1}S}}}{\braket{\braket{S}}},
\label{hs_original}
\end{align}

The mathematical form of \Eq{Noyes} is parallel to that of the Noyes expression for the rate coefficient of bimolecular reactions.\cite{noyes1954treatment}
It is known that the Noyes expression for bimolecular reactions can be expressed as a series expansion using the operator algebraic method.\cite{Kasahara_2021,kim2009rigorous}
As well as in the case of bimolecular reactions, a series expansion for $\hat{\mathcal{P}}\left(s\right)$ (\Eq{hs_original}) can be derived as described below. 
The Laplace transform of \Eq{Mori_identity} is
\begin{align}
\left(s+\mathcal{L}+S\right)^{-1} & =\left(s+\mathcal{L}\right)^{-1} \notag \\
                                  & \quad -\left(s+\mathcal{L}+S\right)^{-1}S\left(s+\mathcal{L}\right)^{-1}. \label{Mori_identity_Laplace}
\end{align}
The following identity is obtained by iteratively utilizing \Eq{Mori_identity_Laplace}.  
\begin{align}
\left(s+\mathcal{L}+S\right)^{-1}S & =-\sum_{n=1}^{\infty}\left(-1\right)^{n}\left[\left(s+\mathcal{L}\right)^{-1}S\right]^{n}. \label{Mori_identity_Laplace_iterative}
\end{align}
By substituting \Eq{Mori_identity_Laplace_iterative} into \Eq{hs_original}, one can obtain 
\begin{align}
\hat{h}\left(s\right) & =-\sum_{n=1}^{\infty}\left(-1\right)^{n}\hat{N}_{n}\left(s\right),
\label{hs_series_expansion}
\end{align}
where $\hat{N}_{n}\left(s\right)$ is a multiple sink correlation function defined as 
\begin{align}
\hat{N}_{n}\left(s\right) & =\dfrac{\braket{\braket{S\left[\left(s+\mathcal{L}\right)^{-1}S\right]^{n}}}}{\braket{\braket{S}}},
\label{Nn}
\end{align}
As described in \Refs{Kasahara_2021}{kim2009rigorous}, the inverse Laplace transform of $\hat{N}_{n}\left(s\right)$, $N_{n}\left(t\right)$, represents the contribution of the permeants repeatedly visiting phase R ($\bm{\Upsilon}$) to the permeability coefficient.  

Kim and Lee derived an alternative type of series expansion that is useful to systematically improve the approximation.\cite{kim2009rigorous}
To utilize their approach, we rewrite \Eq{Noyes} as 
\begin{align}
s\hat{\mathcal{P}}\left(s\right) & =\mathcal{P}_{\mathrm{eq}}\left(1+\dfrac{\hat{h}\left(s\right)}{1-\hat{h}\left(s\right)}\right)^{-1},
\end{align}
Then, by employing the Maclaurin series
\begin{align}
\dfrac{x}{1-x} & =\sum_{n=1}^{\infty}x^{n},
\end{align}
one can obtain
\begin{align}
s\hat{\mathcal{P}}\left(s\right) & =\mathcal{P}_{\mathrm{eq}}\left\{ 1+\sum_{n=1}^{\infty}\left(\hat{h}\left(s\right)\right)^{n}\right\} ^{-1}. \label{sP_h_expansion}
\end{align}
Substituting \Eq{hs_series_expansion} into $\sum_{n=1}^{\infty}(\hat{h}(s))^{n}$ yields
\begin{align}
\sum_{n=1}^{\infty}\left(\hat{h}\left(s\right)\right)^{n} & =\sum_{m=1}^{\infty}\sum_{n=1}^{\infty}\left\{ \left(-1\right)^{n+1}\hat{N}_{n}\left(s\right)\right\} ^{m}\notag\\
 & =\sum_{n=1}^{\infty}\left(-1\right)^{n+1}\hat{Y}_{n}\left(s\right),
\end{align}
where $\{\hat{Y}_{n}\left(s\right)\}$ is defined recursively as
\begin{align}
\hat{Y}_{1}\left(s\right) & =\hat{N}_{1}\left(s\right), \label{Y1}\\
\hat{Y}_{2}\left(s\right) & =\hat{N}_{2}\left(s\right)-\hat{N}_{1}\left(s\right)\hat{Y}_{1}\left(s\right), \label{Y2}\\
\hat{Y}_{3}\left(s\right) & =\hat{N}_{3}\left(s\right)-\hat{N}_{2}\left(s\right)\hat{Y}_{1}\left(s\right)-\hat{N}_{1}\left(s\right)\hat{Y}_{2}\left(s\right),\\
\cdots & \notag\\
\hat{Y}_{n}\left(s\right) & =\hat{N}_{n}\left(s\right)-\sum_{m=1}^{n-1}\hat{N}_{m}\left(s\right)\hat{Y}_{n-m}\left(s\right). \label{Yn}
\end{align}
Thus, \Eq{sP_h_expansion} can be expressed as 
\begin{align}
s\hat{\mathcal{P}}\left(s\right) & =\mathcal{P}_{\mathrm{eq}}\left\{ 1+\sum_{n=1}^{\infty}\left(-1\right)^{n+1}\hat{Y}_{n}\left(s\right)\right\}^{-1}. \label{sP_KL}
\end{align}

Introducing an approximation to $\hat{N}_{n}\left(s\right)$ gives the tractable expression of $\hat{\mathcal{P}}\left(s\right)$ from \Eq{sP_KL}.
%
%
For notational simplicity, we let $\bm{\zeta}_{i}$ be $\left(z_{i},\bm{\Lambda}_{i}\right)$.
Then, $N_{1}\left(t\right)$ is expressed as 
\begin{align}
 & N_{1}\left(t\right) =\dfrac{\braket{\braket{Se^{-\mathcal{L}t}S}}}{\braket{\braket{S}}}\notag\\
 & =\dfrac{\displaystyle \int d\bm{\xi}_{1}\int d\bm{\xi}_{0}\,S\left(\bm{\xi}_{1}\right)S\left(\bm{\xi}_{0}\right)\braket{\braket{\delta\left(\bm{\xi}_{1}-\bm{\zeta}_{i}\left(t\right)\right)\delta\left(\bm{\xi}_{0}-\bm{\zeta}_{i}\right)}}}{\displaystyle \int d\bm{\xi}_{0}\,S\left(\bm{\xi}_{0}\right)\braket{\braket{\delta\left(\bm{\xi}_{0}-\bm{\zeta}_{i}\right)}}},
\label{N1}
\end{align}
where we have used the relationship given by
$S\left(\bm{\zeta}_{i}\right)e^{-\mathcal{L}t} =S\left(\bm{\zeta}_{i}\left(t\right)\right)$.
Substituting Eq. \eqref{sink_definition} into \Eq{N1} yields 
\begin{align}
N_{1}\left(t\right) & =k_{\mathrm{R\to A}}P_{\mathrm{RET}}\left(t\right). \label{N1_Pret}
\end{align}
Here, $P_{\mathrm{RET}}\left(t\right)$ is the returning probability defined as 
\begin{align}
P_{\mathrm{RET}}\left(t\right) & =\dfrac{\displaystyle \int_{\bm{\Upsilon}}d\bm{\xi}_{1}\int_{\bm{\Upsilon}}d\bm{\xi}_{0}\,\braket{\braket{\delta\left(\bm{\xi}_{1}-\bm{\zeta}_{i}\left(t\right)\right)\delta\left(\bm{\xi}_{0}-\bm{\zeta}_{i}\right)}}}{\displaystyle \int_{\bm{\Upsilon}} d\bm{\xi}_{0}\,\braket{\braket{\delta\left(\bm{\xi}_{0}-\bm{\zeta}_{i}\right)}}}.
\label{Pret_app}
\end{align}
$P_{\mathrm{RET}}\left(t\right)$ is the conditional probability of finding a permeant in $\bm{\Upsilon}$ at $t=t$ 
when that molecule was in $\bm{\Upsilon}$ at $t=0$. 
Note that \Eq{Pret_app} is equivalent to \Eq{pret_definition}. 
Let us introduce
\begin{align}
\mathcal{G}^{\left(2\right)}\left(\bm{\xi}_{1},t|\bm{\xi}_{0},0\right) & =\dfrac{\braket{\braket{\delta\left(\bm{\xi}_{1}-\bm{\zeta}_{1}\left(t\right)\right)\delta\left(\bm{\xi}_{0}-\bm{\zeta}_{i}\right)}}}{g_{\mathrm{eq}}\left(\bm{\xi}_{0}\right)}, 
\end{align}
where $g_{\mathrm{eq}}\left(\bm{\xi}\right)$ is the equilibrium distribution function on $\bm{\xi} = \left({\bf r},\bm{\Lambda}\right)$ as defined in \Eq{geq_app}.
$\mathcal{G}^{\left(2\right)}\left(\bm{\xi}_{1},t | \bm{\xi}_{0}, 0\right)d\bm{\xi}_{1}d\bm{\xi}_{0}$ 
is the conditional probability of finding a permeant whose coordinate on $\bm{\xi}$-space is $\bm{\xi}_{1}$ in element $d\bm{\xi}_{1}$ at $t=t$, given that the coordinate was $\bm{\xi}_{0}$ in $d\bm{\xi}_{0}$ at $t=0$.
\Eq{Pret_app} can be rewritten as
\begin{align}
P_{\mathrm{RET}}\left(t\right) & =\dfrac{{\displaystyle \int_{\bm{\Upsilon}}d\bm{\xi}_{1}\int_{\bm{\Upsilon}}d\bm{\xi}_{0}\,\mathcal{G}^{\left(2\right)}\left(\bm{\xi}_{1},t|\bm{\xi}_{0},0\right)g_{\mathrm{eq}}\left(\bm{\xi}_{0}\right)}}{{\displaystyle \int_{\bm{\Upsilon}}d\bm{\xi}_{0}\,}g_{\mathrm{eq}}\left(\bm{\xi}_{0}\right)},
\end{align}
Similar to $N_{1}\left(t\right)$, $N_{n}\left(t\right)$ is expressed from \Eq{Nn} as 
\begin{widetext}
\begin{align}
N_{n}\left(t\right) & =\dfrac{1}{\braket{\braket{S}}}\int_{0}^{t}d\tau_{n-1}\int_{0}^{\tau_{n-1}}d\tau_{n-2}\cdots\int_{0}^{\tau_{2}}d\tau_{1}\,\braket{\braket{Se^{-\mathcal{L}t}Se^{-\mathcal{L}\tau_{n-1}}S\cdots Se^{-\mathcal{L}\tau_{1}}S}}.
\end{align}
From \Eq{sink_definition}, the above equation can be rewritten as 
\begin{align}
\dfrac{N_{n}\left(t\right)}{k^{n}_{\mathrm{R\to A}}} & =\dfrac{{\displaystyle \int_{\bm{\Upsilon}}d\bm{\xi}_{n}\int_{\bm{\Upsilon}}d\bm{\xi}_{0}}\,\Braket{\Braket{\delta\left(\bm{\xi}_{n}-\bm{\zeta}_{i}\left(t\right)\right)\left[{\displaystyle \prod_{k=1}^{n-1}\int_{0}^{\tau_{k+1}} d\tau_{k}\int_{\bm{\Upsilon}}d\bm{\xi}_{k}\,\delta\left(\bm{\xi}_{k}-\bm{\zeta}_{i}\left(\tau_{k}\right)\right)}\right]\delta\left(\bm{\xi}_{0}-\bm{\zeta}_{i}\right)}}}{{\displaystyle \int_{\bm{\Upsilon}}d\bm{\xi}_{0}\,g_{\mathrm{eq}}\left(\bm{\xi}_{0}\right)}}, \label{Nn_t}
\end{align}
where $\tau_{n} = t$.
$N_{n}\left(t\right)$ can be interpreted as the probability for a permeant to repeatedly visit $\bm{\Upsilon}$.
By assuming the Markovianity that a visiting event to $\bm{\Upsilon}$ is independent of the previous events, one can obtain 
\begin{align}
\dfrac{N_{n}\left(t\right)}{k_{\mathrm{R\to A}}^{n}} & \approx\dfrac{1}{{\displaystyle \int_{\bm{\Upsilon}}d\bm{\xi}_{0}\,g_{\mathrm{eq}}\left(\bm{\xi}_{0}\right)}}\int_{0}^{t}d\tau_{n-1}\int_{0}^{\tau_{n-1}}d\tau_{n-2}\cdots\int_{0}^{\tau_{2}}d\tau_{1}\int_{\bm{\Upsilon}}d\bm{\xi}_{n}\cdots\int_{\bm{\Upsilon}}d\bm{\xi}_{0}\,
 \left[\prod_{k=0}^{n-1}\mathcal{G}^{\left(2\right)}\left(\bm{\xi}_{k+1},\tau_{k+1}|\bm{\xi}_{k},\tau_{k}\right)\right]g_{\mathrm{eq}}\left(\bm{\xi}_{0}\right). \label{Nn_Markov}
\end{align}
Note that each visiting event is represented with a $\mathcal{G}^{\left(2\right)}$-function in this approximation.
Since the $\mathcal{G}^{\left(2\right)}$-function depends on time in the form of $\tau_{k+1}-\tau_{k}$, 
\Eq{Nn_Markov} also assumes the local equilibrium within $\bm{\Upsilon}$.  
The integration over $\bm{\xi}_{k}$ ($k=1,2,\cdots, n$) in \Eq{Nn_Markov} 
involves the two successive $\mathcal{G}^{\left(2\right)}$-functions.
Then, we introduce a mean-field type approximation expressed as
\begin{align}
 & \int d\bm{\xi}_{k}\,\mathcal{G}^{\left(2\right)}\left(\bm{\xi}_{k+1},\tau_{k+1}|\bm{\zeta}_{k},\tau_{k}\right)\mathcal{G}^{\left(2\right)}\left(\bm{\xi}_{k},\tau_{k}|\bm{\zeta}_{k-1},\tau_{k-1}\right) \notag \\
 & \approx\dfrac{{\displaystyle \int_{\bm{\Upsilon}}d\bm{\xi}_{k}}\,\mathcal{G}^{\left(2\right)}\left(\bm{\xi}_{k+1},\tau_{k+1}|\bm{\xi}_{k},\tau_{k}\right)g_{\mathrm{eq}}\left(\bm{\xi}_{k}\right)}{{\displaystyle \int_{\bm{\Upsilon}}d\bm{\xi}_{k}\,g_{\mathrm{eq}}\left(\bm{\xi}_{k}\right)}}\int_{\bm{\Upsilon}}d\bm{\xi}_{k}\,\mathcal{G}^{\left(2\right)}\left(\bm{\xi}_{k},\tau_{k}|\bm{\xi}_{k-1},\tau_{k-1}\right). \label{mean_field}
\end{align} 
Adopting this approximation to \Eq{Nn_Markov} yields
\begin{align}
\dfrac{N_{n}\left(t\right)}{k_{\mathrm{R\to A}}^{n}} & = \int_{0}^{t}d\tau_{n-1}\int_{0}^{\tau_{n-1}}d\tau_{n-2}\cdots\int_{0}^{\tau_{2}}d\tau_{1}\,\left[\prod_{k=0}^{n-1}\dfrac{{\displaystyle \int_{\bm{\Upsilon}}d\bm{\xi}_{k+1}\int_{\bm{\Upsilon}}d\bm{\xi}_{k}\,}\mathcal{G}^{\left(2\right)}\left(\bm{\xi}_{k+1},\tau_{k+1}|\bm{\xi}_{k},\tau_{k}\right)g_{\mathrm{eq}}\left(\bm{\xi}_{k}\right)}{{\displaystyle \int_{\bm{\Upsilon}}d\bm{\xi}_{k}\,g_{\mathrm{eq}}\left(\bm{\xi}_{k}\right)}}\right]\notag\\
 & =\int_{0}^{t}d\tau_{n-1}\int_{0}^{\tau_{n-1}}d\tau_{n-2}\cdots\int_{0}^{\tau_{2}}d\tau_{1}\,\dfrac{1}{k_{\mathrm{R\to A}}^{n}}\left[\prod_{k=0}^{n-1}N_{1}\left(\tau_{k+1}-\tau_{k}\right)\right]. \label{Nn_Markov_Mean_field}
\end{align}
\end{widetext}
%
Note that \Eq{Nn_Markov_Mean_field} is equivalent to the Wilemski-Fixman decoupling 
approximation\cite{wilemski1973general} described as 
\begin{align}
\hat{N}_{n}\left(s\right) & \approx\left(\hat{N}_{1}\left(s\right)\right)^{n}. \label{WF_approx}
\end{align}
%
%
Under this approximation, $\{\hat{Y}_{n}\left(s\right)\}$ (\Eq{Yn})  reduces to zero for $n\geq 2$.  
Accordingly, 
one can obtain the tractable expression of $\hat{\mathcal{P}}\left(s\right)$ from Eqs. \eqref{Y1} and \eqref{sP_KL} as
\begin{align}
s\hat{\mathcal{P}}\left(s\right) & =\mathcal{P}_{\mathrm{eq}}\left(1+k_{\mathrm{R\to A}}\hat{P}_{\mathrm{RET}}\left(s\right)\right)^{-1},
\end{align}
where $\hat{P}_{\mathrm{RET}}\left(s\right)$ is the Laplace transform of $P_{\mathrm{RET}}\left(t\right)$.
The final value theorem of the Laplace transform, $s\hat{\mathcal{P}}\left(s\right) \xrightarrow[]{s\to 0} \mathcal{P}_{\mathrm{ss}}$, gives
\begin{align}
 \mathcal{P}_{\mathrm{ss}} = \mathcal{P}_{\mathrm{eq}}\left(1+k_{\mathrm{R\to A}}\int_{0}^{\infty}dt\,P_{\mathrm{RET}}\left(t\right)\right)^{-1}. \label{RP_Markov} 
\end{align}
According to the theoretical analysis using the Smoluchowski equation, 
the decoupling approximation (\Eq{WF_approx}) becomes accurate with sufficiently narrow $\bm{\Upsilon}$ 
when the underlying dynamics obeys the Smoluchowski equation.\cite{kim2009rigorous}
On the other hand, the estimated value of kinetic properties such as binding rate constant 
using too narrow $\bm{\Upsilon}$ is found to be strongly affected by the variation in $\bm{\Upsilon}$.\cite{kasahara2023elucidating}
This suggests that the inertial effect of molecular motions that is present in realistic systems 
might violate the local-equilibrium condition within $\bm{\Upsilon}$ assumed in \Eq{Nn_Markov} 
in the case of too narrow $\bm{\Upsilon}$.
Accordingly,
the appropriate $\bm{\Upsilon}$ should be determined so that the kinetic properties are hardly affected by the variation in $\bm{\Upsilon}$. 
\section{Improved expression of $\mathcal{P}_{\mathrm{ss}}$\label{sec:Improved_Pss}}
Equation \eqref{RP_Markov} is derived by assuming the Markovianity that the repeated vising events of a permeant to $\bm{\Upsilon}$
are uncorrelated with each other.
The perturbative expansion of $\hat{\mathcal{P}}\left(s\right)$ (\Eq{sP_KL}) 
provides us a route to include the non-Markovianity partly by treating $\hat{Y}_{n}$-terms up to $n=2$ as 
\begin{align}
s\hat{\mathcal{P}}\left(s\right) & =\mathcal{P}_{\mathrm{eq}}\left(1+k_{\mathrm{R\to A}}\hat{P}_{\mathrm{RET}}\left(s\right)-\hat{Y}_{2}\left(s\right)\right)^{-1}, \label{sP_KL_2nd_order}
\end{align}
where we have used \Eqs{Y1}{N1_Pret}.
By substituting \Eqs{Y1}{N1_Pret} into \Eq{Y2}, $\hat{Y}_{2}\left(s\right)$ is expressed as
\begin{align}
\hat{Y}_{2}\left(s\right) & =\hat{N}_{2}\left(s\right)-k_{\mathrm{R\to A}}^{2}\left(\hat{P}_{\mathrm{RET}}\left(s\right)\right)^{2}. \label{Y2_Pret}
\end{align} 
%
From \Eq{Nn_t}, $N_{2}\left(t\right)$ is given by
\begin{align}
N_{2}\left(t\right) & =k_{\mathrm{R\to A}}^{2}\int_{0}^{t}d\tau\, P_{\mathrm{RET}}^{\left(2\right)}\left(t,\tau\right), \label{N2_Pret_2nd_order}
\end{align}
where
\begin{align}
 & P_{\mathrm{RET}}^{\left(2\right)}\left(t,\tau\right)=\dfrac{1}{{\displaystyle \int_{\bm{\Upsilon}}d\bm{\zeta}_{0}\, g_{\mathrm{eq}}\left(\bm{\xi}_{0}\right)}}\int_{\bm{\Upsilon}}d\bm{\zeta}_{2}\int_{\bm{\Upsilon}}d\bm{\zeta}_{1} \int_{\bm{\Upsilon}}d\bm{\zeta}_{0}\notag\\
 & \times\braket{\braket{\delta\left(\bm{\zeta}_{2}-\bm{\zeta}_{i}\left(t\right)\right)\delta\left(\bm{\zeta}_{1}-\bm{\zeta}_{i}\left(\tau\right)\right)\delta\left(\bm{\zeta}_{0}-\bm{\zeta}_{i}\right)}}.
\end{align}
$P_{\mathrm{RET}}^{\left(2\right)}\left(t,\tau\right)$ is the conditional probability of finding a permeant in $\bm{\Upsilon}$ at $t=\tau$ and $t$ when that molecule was in $\bm{\Upsilon}$ at $t=0$.
Therefore, adopting the final value theorem to \Eq{sP_KL_2nd_order} with \Eqs{Y2_Pret}{N2_Pret_2nd_order} gives the following expression of $\mathcal{P}_{\mathrm{ss}}$.
\begin{widetext}
\begin{align}
  \mathcal{P}_{\mathrm{ss}} &=\mathcal{P}_{\mathrm{eq}}\left[1+k_{\mathrm{R\to A}}\int_{0}^{\infty}dt\,P_{\mathrm{RET}}\left(t\right)
 -k_{\mathrm{R\to A}}^{2}\left\{ \int_{0}^{\infty}dt\int_{0}^{\infty}d\tau\,P_{\mathrm{RET}}^{\left(2\right)}\left(t,\tau\right) 
 -\left(\int_{0}^{\infty}dt\,P_{\mathrm{RET}}\left(t\right)\right)^{2}\right\} \right]^{-1}. \label{Improved_Pss}
\end{align}
\end{widetext}
\bibliographystyle{apsrev4-1}
\bibliography{rpmemb}
\clearpage
\widetext

\def\thesection{S\arabic{section}}
\setcounter{section}{0}
\renewcommand{\theequation}{S\arabic{equation}}
\setcounter{equation}{0}
\renewcommand{\thefigure}{S\arabic{figure}}
\setcounter{figure}{0}
\renewcommand{\thetable}{S\arabic{table}}
\setcounter{table}{0}
\renewcommand{\thepage}{S\arabic{page}}
\setcounter{page}{0}

\begin{center}
\Large \bf Supplement for ``A methodology of quantifying membrane permeability based on returning probability theory and molecular dynamics simulation''
\end{center}
\section{Simulation protocols for equilibrating membrane systems}
\begin{figure}[h]
  \centering
  \includegraphics[width=1.0\linewidth]{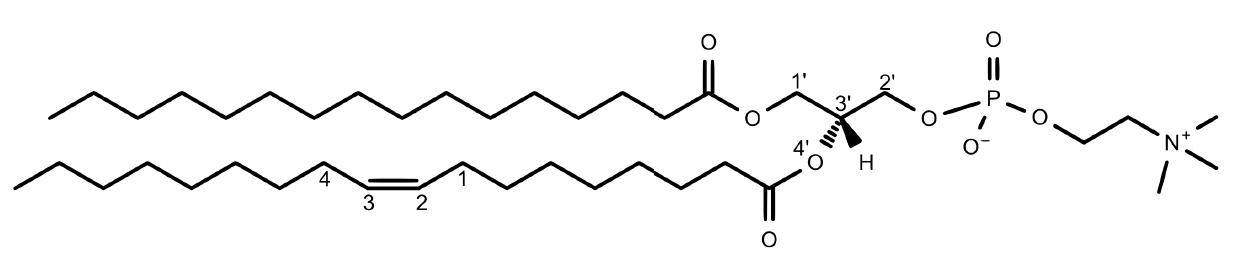}
  \caption{Chemical structure of 1-palmitoyl-2-oleoyl-\textit{sn}-glycero-3-phosphocholin (POPC). The atoms related with the torsion harmonic potentials imposed during the equilibration are labeled.\label{fig:POPC}} 
\end{figure}

%
In Table S1, POSI, FB, and TORS denote the positional harmonic restraints, flat-bottom, and torsion harmonic restrains, respectively.
POSI is defined as
\begin{align}
  \text{POSI}(\mathrm{M},k)= \sum_{i\in \mathrm{M}}k\left(z_i - z_{i,0}\right)^2,\\
  \text{POSI}(\mathrm{U},k)= k\left(z_{\mathrm{U}} - z_{\mathrm{U},0}\right)^2.
\end{align}
The arguments M and U in POSI respectively mean that the positional restraints are imposed on the $z$-coordinate of the phosphorus atoms in the membrane and on the $z$-component of the center of mass (CoM) for the permeant solute molecule.
The CoM is calculated from the heavy atoms in the permeant, and 
$z_{i}$ is the $z$-coordinate of the $i$th phosphorus atom in the membrane with its reference position $z_{i,0}$.
$z_{\mathrm{U}}$ is the $z$-coordinate of CoM for a permeant molecule with its initial position $z_{\mathrm{U},0}$.
$\mathrm{POSI}\left(\mathrm{U},k\right)$ is imposed only for the membrane systems containing one permeant molecule, and $z_{\mathrm{U},0}$ is set to 0. 
The unit of $k$ is $\mathrm{kcal~mol^{-1}~\AA^{-2}}$. 
The definition of FB is
\begin{align}
  \mathrm{FB}(z_1,z_2,k) &=
  \begin{cases}
    k(z_\mathrm{U}-z_1)^2, & z_\mathrm{U}\leq z_1, \\
    0,                     & z_1 < z_\mathrm{U}\leq z_2, \\
    k(z_\mathrm{U}-z_2)^2, & z_2\leq z_\mathrm{U}, \\
  \end{cases}
\end{align}
where the units of $z_{1}$, $z_{2}$ and $k$ are $\mathrm{\AA}$, $\mathrm{\AA}$, and $\mathrm{kcal~mol^{-1}~\mathrm{\AA}^{-2}}$, respectively.
$\mathrm{TORS}$ is defined as
\begin{align}
  \mathrm{TORS}(\mathrm{X},k) = k(\phi - \phi_0)^2,
\end{align}
where $\phi$ is the proper torsion angle for the double bond with cis-form in the acyl chain ($\mathrm{X}=\mathrm{A}$) 
and the improper torsion (inversion) angle for the glycerol group ($\mathrm{X}=\mathrm{G}$).
%
The atoms used for defining the torsion angles are labeled in \Fig{POPC}. 
The torsion angles are composed of atoms $(1,2,3,4)$ for $\mathrm{X}=\mathrm{A}$ and of $(1^{\prime},2^{\prime},3^{\prime},4^{\prime})$ for $\mathrm{X}=\mathrm{G}$.
The values of $\phi_{0}$ are 0 and 120 degrees for $\mathrm{X}=\mathrm{A}$ and $\mathrm{G}$, respectively.
The unit of $k$ is $\mathrm{kcal~mol^{-1}~degree^{-2}}$. 
\newpage
%
%
%
\begin{table}[h]
\centering
\caption{Simulation protocols for equilibrating membrane systems. The velocity Verlet (VVER) integrator is used at all the steps, and $\Delta t$  means the time interval used in VVER integrator. 
As for the systems containing one permeant molecule, $\mathrm{POSI}\left(\mathrm{U},k=1.0\right)$) is imposed at all the steps. \label{equil_RP}}
\begin{tabular}{ccccc}
\hline 
\hline
Step & Simul. length & $\Delta t$ & Ensemble & Restraints\tabularnewline
\hline 
1 & 0.125 ns & 1 fs & NVT & $\mathrm{POSI}\left(\mathrm{M},k=2.5\right)$, $\mathrm{TORS}\left(\mathrm{A},k=250\right)$,
$\mathrm{TORS}\left(\mathrm{G},k=250\right)$ \tabularnewline
2 & 0.125 ns & 1 fs & NVT & $\mathrm{POSI}\left(\mathrm{M},k=2.5\right)$, $\mathrm{TORS}\left(\mathrm{A},k=100\right)$,
$\mathrm{TORS}\left(\mathrm{G},k=250\right)$\tabularnewline
3 & 0.125 ns & 1 fs & NPT & $\mathrm{POSI}\left(\mathrm{M},k=1\right)$, $\mathrm{TORS}\left(\mathrm{A},k=50\right)$,
$\mathrm{TORS}\left(\mathrm{G},k=50\right)$\tabularnewline
4 & 0.5 ns & 2 fs & NPT & $\mathrm{POSI}\left(\mathrm{M},k=0.5\right)$, $\mathrm{TORS}\left(\mathrm{A},k=50\right)$,
$\mathrm{TORS}\left(\mathrm{G},k=50\right)$\tabularnewline
5 & 0.5 ns & 2 fs & NPT & $\mathrm{POSI}\left(\mathrm{M},k=0.1\right)$, $\mathrm{TORS}\left(\mathrm{A},k=25\right)$,
$\mathrm{TORS}\left(\mathrm{G},k=25\right)$\tabularnewline
6 & 0.5 ns & 2 fs & NPT & \tabularnewline
\hline 
\end{tabular}
%
%
    
\end{table}
\newpage
\section{Scheme of computing $\Delta \Delta \mu$}
\begin{figure}[h]
  \centering
  \includegraphics[width=0.9\linewidth]{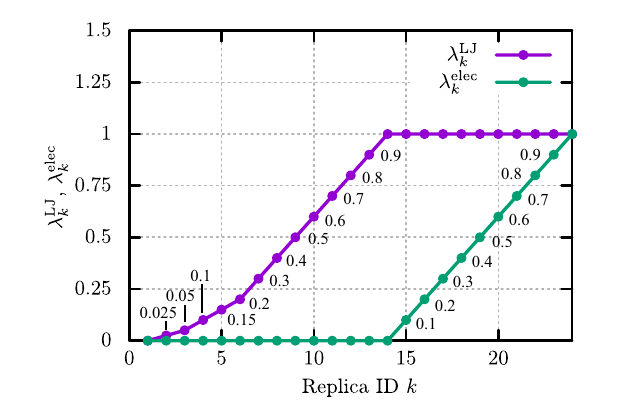}
  \caption{Coupling parameters for the BAR/H-REMD simulations, $\lambda_{k}^{\mathrm{LJ}}$ and $\lambda_{k}^{\mathrm{elec}}$, 
           along replica ID, $k$. \label{fig:lambda}}
\end{figure}

In this section, we describe the scheme of computing
\begin{align}
 \Delta \Delta \mu = \Delta \mu_{\mathcal{S}} - \Delta \mu_{\mathrm{D}}, \label{ddmu} 
\end{align}
using the Bennett acceptance ratio (BAR) method\cite{bennett1976efficient} implemented in GENESIS.\cite{oshima2022modified,matsunaga2022use, jung2015genesis,kobayashi2017genesis,jung2021new} 
Here, $\Delta \mu_{\mathcal{S}}$ and $\Delta \mu_{\mathrm{D}}$ are the solvation free energies associated with the 
solvation processes of the solute in state $\mathcal{S}$ and phase $\mathrm{D}$, respectively.
The free energy perturbation (FEP) method enables us to  compute the free energy difference between two different states
of interest by introducing the set of replicas connecting the two states of interest. 
Let us define the potential function for the $k$th replica ($k=1,\cdots, N_{\mathrm{rep}}$) as  
\begin{align}
\mathcal{V}\left(\bm{\lambda}_{k}\right) & =\lambda_{k}^{\mathrm{LJ}}\left[U_{\mathrm{U}}^{\mathrm{LJ}}\left(\lambda_{k}^{\mathrm{LJ}}\right)+U_{\mathrm{UV}}^{\mathrm{LJ}}\left(\lambda_{k}^{\mathrm{LJ}}\right)\right] \notag \\
 & \quad+\lambda_{k}^{\mathrm{elec}}\left[U_{\mathrm{U}}^{\mathrm{elec}}\left(\lambda_{k}^{\mathrm{elec}}\right)+U_{\mathrm{UV}}^{\mathrm{elec}}\left(\lambda_{k}^{\mathrm{elec}}\right)\right] \notag \\
 & \quad+U_{\mathrm{U}}^{\mathrm{bonded}}+U_{\mathrm{V}}, 
\end{align}
where $\lambda_{k}^{\mathrm{LJ}}$ and $\lambda_{k}^{\mathrm{elec}}$ are the coupling parameters for the Lennard-Jones interaction and electrostatic interaction, respectively, and $\bm{\lambda}_{k}$ is defined as $\bm{\lambda}_{k} = \left(\lambda_{k}^{\mathrm{LJ}}, \lambda_{k}^{\mathrm{elec}}\right)$.
$U_{\mathrm{U}}^{\mathrm{bonded}}$ is the bonded part of the intramolecular potential for the 
solute composed of the bond-stretch, bending, and torsion potentials.
$U_{\mathrm{V}}$ is the total potential of the solvents.
$U_{\mathrm{U}}^{\mathrm{LJ}}\left(\lambda_{k}^{\mathrm{LJ}}\right)$ and $U_{\mathrm{U}}^{\mathrm{elec}}\left(\lambda_{k}^{\mathrm{elec}}\right)$ are the soft-core LJ and soft-core electrostatic potentials for the solute, respectively, 
and $U_{\mathrm{UV}}^{\mathrm{LJ}}\left(\lambda_{k}^{\mathrm{LJ}}\right)$ and $U_{\mathrm{UV}}^{\mathrm{elec}}\left(\lambda_{k}^{\mathrm{elec}}\right)$ respectively mean the soft-core LJ and soft-core electrostatic potentials between the solute and solvents.
$U_{X}^{\mathrm{LJ}}\left(\lambda_{k}^{\mathrm{LJ}}\right)$ and $U_{X}^{\mathrm{elec}}\left(\lambda_{k}^{\mathrm{elec}}\right)$ ($X=\mathrm{U}~\mathrm{or}~\mathrm{UV}$) are respectively defined as\cite{zacharias1994separation,steinbrecher2011soft} 
\begin{align}
U_{X}^{\mathrm{LJ}}\left(\lambda_{k}^{\mathrm{LJ}}\right) & =\sum_{\left(i,j\right)\in\mathcal{N}_{X}}4\varepsilon_{ij}\left[\left(\dfrac{\sigma_{ij}^{2}}{r_{ij}^{2}+\alpha_{\mathrm{sc}}\left(1-\lambda_{k}^{\mathrm{LJ}}\right)}\right)^{6}-\left(\dfrac{\sigma_{ij}^{2}}{r_{ij}^{2}+\alpha_{\mathrm{sc}}\left(1-\lambda_{k}^{\mathrm{LJ}}\right)}\right)^{3}\right], \label{scLJ}\\
U_{X}^{\mathrm{elec}}\left(\lambda_{k}^{\mathrm{elec}}\right) & =\sum_{\left(i,j\right)\in\mathcal{N}_{X}}\dfrac{q_{i}q_{j}\mathrm{erfc}\left(\kappa^{-1}\sqrt{r_{ij}^{2}+\beta_{\mathrm{sc}}\left(1-\lambda_{k}^{\mathrm{elec}}\right)}\right)}{\sqrt{r_{ij}^{2}+\beta_{\mathrm{sc}}\left(1-\lambda_{k}^{\mathrm{elec}}\right)}} \notag \\
 & \quad+\left(\text{SPME reciprocal and self terms}\right). \label{scELEC}
\end{align}
Here, $\mathcal{N}_{X}$ is the set of pairs of atoms involved in $X$, 
and $r_{ij}$ and $\left(\sigma_{ij}, \varepsilon_{ij} \right)$ are the interatomic distance and Lennard-Jones parameter between the $i$th and $j$th atoms, respectively, 
and $q_{i}$ is the point charge on the $i$th atom. 
$\kappa$ is the screening parameter for the smooth particle-mesh Ewald (SPME) method.\cite{essmann1995smooth}
$\alpha_{\mathrm{sc}}$ and $\beta_{\mathrm{sc}}$ are the soft-core parameters for the LJ and electrostatic interactions, respectively.
In the present study, the values of $\alpha_{\mathrm{sc}}$ and $\beta_{\mathrm{sc}}$ were set to $5~\mathrm{\AA}^{2}$ and $0.5~\mathrm{\AA}^{2}$, respectively.
The number of replicas, $N_{\mathrm{rep}}$, was 24, 
and we set $\bm{\lambda}_{0}$ and $\bm{\lambda}_{N_{\mathrm{rep}}}$ to $\left(0,0\right)$ and $\left(1,1\right)$, respectively.
The values of $\bm{\lambda}_{k}$ are plotted against the replica ID in \Fig{lambda}. 
Note that $\bm{\lambda}_{0}$ and $\bm{\lambda}_{1}$ respectively correspond to 
\begin{align}
\begin{cases}
\lambda^{\mathrm{LJ}}U_{X}^{\mathrm{LJ}}\left(\lambda^{\mathrm{LJ}}\right)\xrightarrow{\lambda^{\mathrm{LJ}}\to0}0,\\
\lambda^{\mathrm{elec}}U_{X}^{\mathrm{elec}}\left(\lambda^{\mathrm{elec}}\right)\xrightarrow{\lambda^{\mathrm{elec}}\to0}0,
\end{cases}
\end{align}
and
\begin{align}
\begin{cases}
\lambda^{\mathrm{LJ}}U_{X}^{\mathrm{LJ}}\left(\lambda^{\mathrm{LJ}}\right)\xrightarrow{\lambda^{\mathrm{LJ}}\to1}\left(\text{original LJ potential}\right),\\
\lambda^{\mathrm{elec}}U_{X}^{\mathrm{elec}}\left(\lambda^{\mathrm{elec}}\right)\xrightarrow{\lambda^{\mathrm{elec}}\to1}\left(\text{original SPME electrostatic potential}\right).
\end{cases} 
\end{align}
Thus, the free energy change defined as 
\begin{align}
\Delta G_{1,A} & =- \dfrac{1}{\beta}\log\braket{\exp\left[-\beta\left(\mathcal{V}(\bm{\lambda}_{N_{\mathrm{rep}}})-\mathcal{V}(\bm{\lambda}_{0})\right)\right]}_{\bm{\lambda}_{0},A}
\end{align}
is associated with the appearance of the nonbonded interaction within the solute 
and that between the solute and solvents at state $A$.
Here, $\braket{\cdots}_{\bm{\lambda}_{0}, A}$ denotes the ensemble average of the system governed by $\mathcal{V}\left(\bm{\lambda}_{0}\right)$ at state $A$. 
By introducing the free energy change 
associated with the appearance of the nonbonded interaction within the solute in the gas phase as
\begin{align}
\Delta G_{2} & =-\dfrac{1}{\beta}\log\dfrac{{\displaystyle \int d{\bf x}_{\mathrm{U}}}\,e^{-\beta U_{\mathrm{U}}}}{{\displaystyle \int d{\bf x}_{\mathrm{U}}\,e^{-\beta U_{\mathrm{U}}^{\mathrm{bonded}}}}}, \label{dG2}
\end{align}
the solvation free energy for state $A$ can be expressed as
\begin{align}
\Delta\mu_{A} & =\Delta G_{1,A}-\Delta G_{2}.
\end{align}
Here, $U_{\mathrm{U}}$ is the intramolecular potential of the solute defined as 
$U_{\mathrm{U}} = U_{\mathrm{U}}^{\mathrm{LJ}}\left(1\right) + U_{\mathrm{U}}^{\mathrm{elec}}\left(1\right) + U_{\mathrm{U}}^{\mathrm{bonded}}$.
From the above expression, 
\Eq{ddmu} can be rewritten as 
\begin{align}
\Delta\Delta\mu & =\Delta G_{1,\mathcal{S}}-\Delta G_{1,\mathrm{D}}.
\end{align}
We computed $\Delta G_{1,\mathcal{S}}$ and $\Delta G_{1,\mathrm{D}}$ by performing the BAR method with the Hamiltonian replica exchange MD (BAR/H-REMD) simulations.\cite{jiang2010free}
\newpage
\section{Permeability coefficient corresponding to the arrival of a permeant at the metastable state}
We introduce the permeability coefficient corresponding to the arrival of a permeant at the metastable (M) state 
within the membrane from the reactive (R) phase, $\mathcal{P}_{\mathrm{ss}}^{\prime}$, defined as 
\begin{align}
\mathcal{P}_{\mathrm{ss}}^{\prime} & =k_{\mathrm{R\to M}}K^{*}\left(1+k_{\mathrm{R\to M}}\int_{0}^{\infty}dt\,P_{\mathrm{RET}}\left(t\right)\right)^{-1}. \label{Pss_prime}
\end{align}
Here, $P_{\mathrm{RET}}\left(t\right)$ and $K^*$ are the returning probability and the equilibrium constant 
for phase R ($\bm{\Upsilon}=\left\{0\leq z \leq z_{\mathrm{R}}\right\}$) defined in Eqs. (14) and (21), respectively.
$k_{\mathrm{R\to M}}$ is the rate constant for the transition from phase R to state M.
State M is defined as $z = 15~\mathrm{\AA}$, which corresponds to the local minimum 
of the potentials of mean force ($\Delta G\left(z\right)$). 
Similar to the computation of $k_{\mathrm{R\to A}}$, we determine $k_{\mathrm{R\to M}}$ using Eq. (46),  
with the entry of the permeant into region $z \leq -15~\mathrm{\AA}$ considered as the transition event.
Note that the transition is defined by the entry into $z \leq -25~\mathrm{\AA}$ for $k_{\mathrm{R\to A}}$. 
The $z_{\mathrm{R}}$-dependency of $\mathcal{P}_{\mathrm{ss}}^{\prime}$ is illustrated in \Fig{Perm_loose}.
It is seen that the $z_{\mathrm{R}}$-dependency of $\mathcal{P}_{\mathrm{ss}}^{\prime}$ 
closely resembles that of $\mathcal{P}_{\mathrm{ss}}$ for both the ethanol and methylamine systems (Fig. 6).
Thus, the presence of the multiple states involved in the crossing process ($\mathrm{R\to A}$) 
is not relevant with the $z_{\mathrm{R}}$-dependency of $\mathcal{P}_{\mathrm{ss}}$. 
\newpage
\section{supplementary figures}

%
%
%

\vspace*{3cm}
\newpage
\begin{figure}[h]
  \centering
  \includegraphics[width=1.0\linewidth]{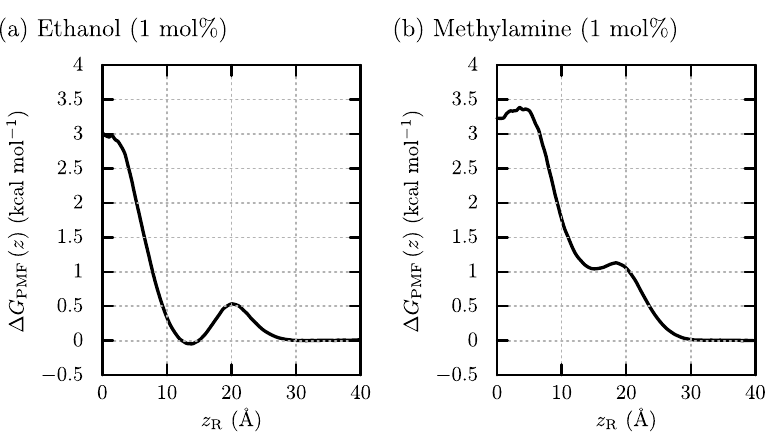}
  \caption{Potentials of mean force (PMF), $\Delta G_{\mathrm{PMF}}\left(z\right)$, obtained from the simulations at the permeant concentrations of 1 mol\% for the transition-based counting (TBC) method. (a) 1 mol\% ethanol, and (b) 1 mol\% methylamine.}
  \label{fig:PMF_conting}
\end{figure}
\begin{figure}[h]
  \centering
  \includegraphics[width=1.0\linewidth]{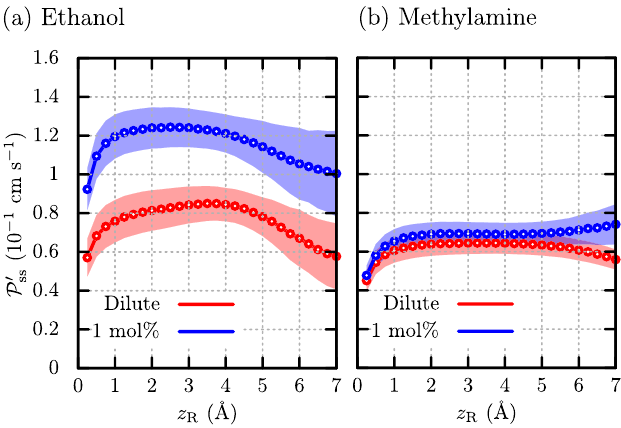}
  \caption{$z_{\mathrm{R}}$-dependency of the permeability coefficients corresponding to the arrival of a permeant at the metastable state ($z \sim${}$-15~\mathrm{\AA}$), $\mathcal{P}_{\mathrm{ss}}^{\prime}$ (\Eq{Pss_prime}), for (a) ethanol and (b) methylamine. The colored regions indicate the statistical uncertainty (standard error). \label{fig:Perm_loose}}
\end{figure}
\newpage
%
%
\begin{figure}[h]
  \centering
  \includegraphics[width=0.9\linewidth]{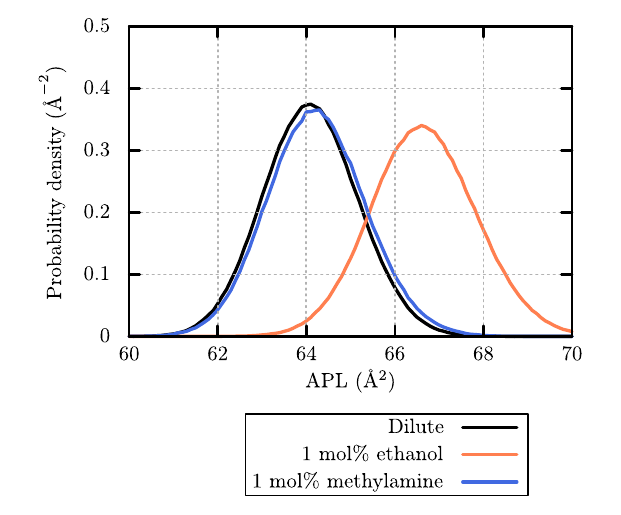}
  \caption{Probability densities of area per lipid (APL) for the dilute ethanol, 1 mol\% ethanol, and 1 mol\% methylamine systems.
           The MD trajectories prepared for determining $k_{\mathrm{R\to A}}$ (Sec. IIIC) are used to compute the probability densities.\label{fig:APL_distribution}}
\end{figure}
\clearpage
%
%

\end{document}